\documentclass[11pt,preprint,letterpaper,xdvi]{aastex}
\usepackage{amsmath} 
\usepackage{wrapfig} 
\usepackage{array}  
\usepackage{brief_bib}
		
\renewcommand{\refname}{} 
\renewcommand{\bibname}{}

\newcommand\bvec{{\bf B}}

\def\referee#1{{#1}}
\def\refereetwo#1{{#1}}

\begin{document}

\title{\referee{On Anisotropy in Expansion of Magnetic Flux Tubes in the Solar Corona}}
\author{A.~Malanushenko$^{1,2}$, C.~J.~Schrijver$^2$}
\affil{$^1$Department of Physics, Montana State University, Bozeman, MT, USA\\
       $^2$Lockheed Martin Advanced Technology Center, Palo Alto, CA, USA}

\begin{abstract}
Most 1d hydrodynamic models of plasma confined to magnetic flux tubes assume circular cross-section of these tubes. We use potential field models to show that flux tubes in circumstances relevant to the solar corona do not in general maintain the same cross-sectional shape through their length and therefore the assumption of a circular cross-section is rarely true. We support our hypothesis with mathematical reasoning and numeric experiments. We demonstrate that lifting this assumption in realistic non-circular loops make apparent expansion of magnetic flux tubes consistent with that of observed coronal loops. We propose that in a bundle of ribbon-like loops those that are viewed along the wide direction would stand out against those that are viewed across the wide direction, due to the difference in their column depths. That would impose a bias towards selecting loops that appear not to be expanding seen projected in the plane of sky. An implication of this selection bias is that the preferentially selected non-circular loops would appear to have increased pressure scale height even if they are resolved by current instruments.
\end{abstract}

\section{Introduction}\label{sec_intro}

Coronal loops appear like thin emitting strands in the solar atmosphere. They appear to follow lines of the magnetic field in the corona and are usually considered to be a manifestation of hot and dense plasma confined within magnetic flux tubes. Because of the frozen-in condition, plasma cannot leave a flux tube through its sides so a local enhancement of density could only get redistributed along a flux tube it is embedded in. Emission measure is proportional to the square of density so \refereetwo{}flux tubes with denser plasma produce more emission. As thermal conductivity is much greater along the field than across it, a localized heating event heats the flux tube it is embedded \refereetwo{in} but not its surroundings\refereetwo{. S}hould the heating be sufficient for plasma to emit in an observable wavelength and should the plasma in a given flux tube be denser than its surroundings, imaging instruments would detect a coronal loop. 

Coronal loops have been observed for several decades, yet some of their properties remain a mystery. For example it is recognized that the width of coronal loops varies only weakly with height above the surface \citep[e.g.,][and references therein]{Klimchuk2000}. At the same time if the emitting plasma is indeed confined to a magnetic flux tube, it should expand with height as the strength of the field drops and the flux is constant along the tube. 

Various theories were proposed to explain this phenomenon. For example, it \citet{McClymont1994} showed that twisted flux tubes expand somewhat less with height than if they were untwisted, but they have also found that for currents typical for active region fields the reported effect is not enough to reach agreement with observations. In another study, \citet{Klimchuk2000b} examined a twisted flux tube and concluded that the presence of electric currents is capable of enforcing circularity in cross-section of the tube. However, for any amount of twist they injected, only one of the shells of the tube was circular at the apex (and all other shells were oblate, as follows from comparing Figures 6 and 9 in their study). In particular, they found that the core of the tube was circular in the apex for a twist value slightly below $2\pi$ (and for this value of twist the surrounding shells were oblate at the apex). Klimchuk et al. speculated that for increased amount of twist the entire flux tube might become increasingly more circular. Their particular simulation was limited to twist values of $\approx 2\pi$, but they mention \citet{Mikic1990b} to justify that twist of up to $5\pi$ could be introduced to a flux tube without loss of equilibrium. However, flux tubes simulated in a later study by \citet{Amari1999c} became kink unstable at twist $\approx 2.7\pi$, which may be below the amount that, according to Klimchuk et al, needs to be injected to enforce circularity of the entire flux tube. 

It was also shown that magnetic flux tubes expand less near separators \citep{Plowman2009}. This finding compares favorably with the observations that cool loop fans are associated with quasi-separatrix layers \citep{Schrijver2010}; however, solid evidence that loops mainly form at the separators are yet to be collected.

\referee{A hypothesis \refereetwo{gaining increasing popularity states} that loops are composed of thin unresolved strands \citep{DeForest2007}. The author argues that the lack of observed expansion of loops with height could be explained in terms of the interaction between resolution, geometric cross-section and a diffuse background.} This is a simple and elegant \referee{theory} which also explains several other puzzling aspects of coronal loops. Among them is the fact that as resolution of our instruments increases, progressively finer strands are reported in what was previously thought of as monolithic structures. Another aspect explained by the thin strand theory is the apparent increase in pressure scale height by several orders of magnitude in \referee{some EUV} loops compared to theoretical predictions \citep[e.g.,][and references therein]{Warren2003b, Winebarger2003}. \referee{\refereetwo{In} a simple model, \citet{DeForest2007} shows that \refereetwo{} resolved and \refereetwo{} unresolved expanding strands would fade with height according to different power laws. We discuss this model in \refereetwo{further detail later}in the text.} Also, an ensemble of thin strands was shown to be brighter than predicted by static loops and appear to be isothermal in filter ratio measurements, which is consistent with observations \citep{Warren2002}. \referee{However, \citet{Fuentes2008} argued that it is possible to distinguish between unresolved but expanding and not expanding but resolved strands on the observations.} 

Every single one of the models we mentioned, as well as, \referee{to our knowledge}, almost all other models of an isolated flux tube\footnote{With exception of several works on loop oscillations. E.g., \citet{Ruderman2003} analyzed flux tubes with constant but elliptic cross-section and showed that they have two sets of eigenfrequencies. The resonant damping properties of oscillations in this case depend on the aspect ratio of the ellipse; however, they concluded that the damping times stay within the order of magnitude of those for a tube with circular cross-section.}, make an assumption that the flux tube is circular in cross-section \referee{\citep[or that the cross-section is constant and therefore its shape is irrelevant, e.g.,][]{Rosner1978}}. This is mainly caused by the fact that it substantially simplifies the calculations. In the absence of definite information about cross-sectional shape of the loops \referee{this is a reasonable simplification to make}. 

Observational evidence that coronal loops are circular in cross-section is extremely difficult to \refereetwo{obtain} and there are not many studies which address this question. This is caused by the fact that the corona is optically thin and what is observed is the superposition of a complicated background and multiple loops (or fragments of loops) along the line of sight. Among the most frequently cited is the study of \citet{Klimchuk2000}, who concluded \refereetwo{} that the loop profiles are typically simple and single-peaked and the width variations along the length of the loops are small. He argued that this supports the hypothesis that loops are circular in cross-section. 

\referee{There are also several studies which provide evidences of the opposite.} For example, \citet{Wang1998} examined a flux tube of each of several model fields \citep{LowLou1990} and concluded that in the presence of currents, flux tubes tend to expand in an anisotropic manner (however they found that the anisotropy is smaller for strongly sheared fields). \citet{Fuentes2006} also mentioned that flux tubes could be strongly non-circular in cross-section. They; however, argued that on a large statistical sample of loops the expansions seen in the ``wide'' and in the ``narrow'' directions would average out. 

More evidence that loops are non-circular in cross-section could be drawn from 3D MHD models of coronal loops. For example, \citet{Gudiksen2005b} examined the shapes of the bases of flux tubes which were round at the apex and confirmed their theoretical argument that the expansion of flux tubes is not isotropic in all directions\refereetwo{. Also, they} mentioned that in their simulation the flux tubes initially round at the apex become ``wrinkled'' at the lower boundary. The studies of \citet{Mok2008} and \citet{Peter2012} do not address this issue directly; however a visual examination of different renderings in their simulations, namely the panels of Fig. 3 of Mok et. al. and Figs 1 and 5 in Peter \& Bingert, \refereetwo{suggests} that the loops in these models also appear to have different width when viewed from different directions. 
 
\begin{figure}[!hc] 
  \begin{center} 
   \begin{tabular}{cc}
   & \\
   & \\
     \includegraphics[height=4cm]{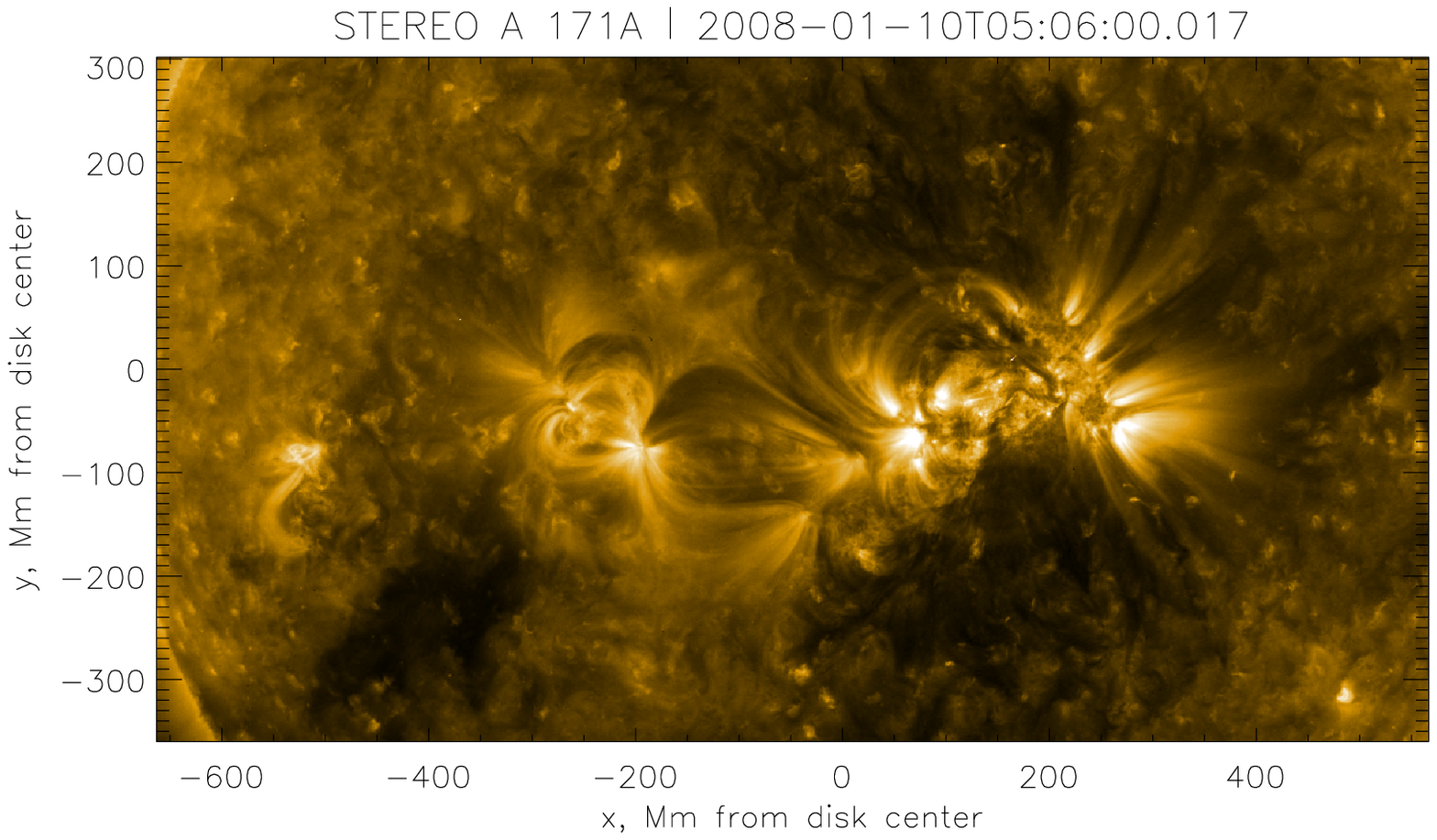} & \hspace{1cm}\includegraphics[height=4cm]{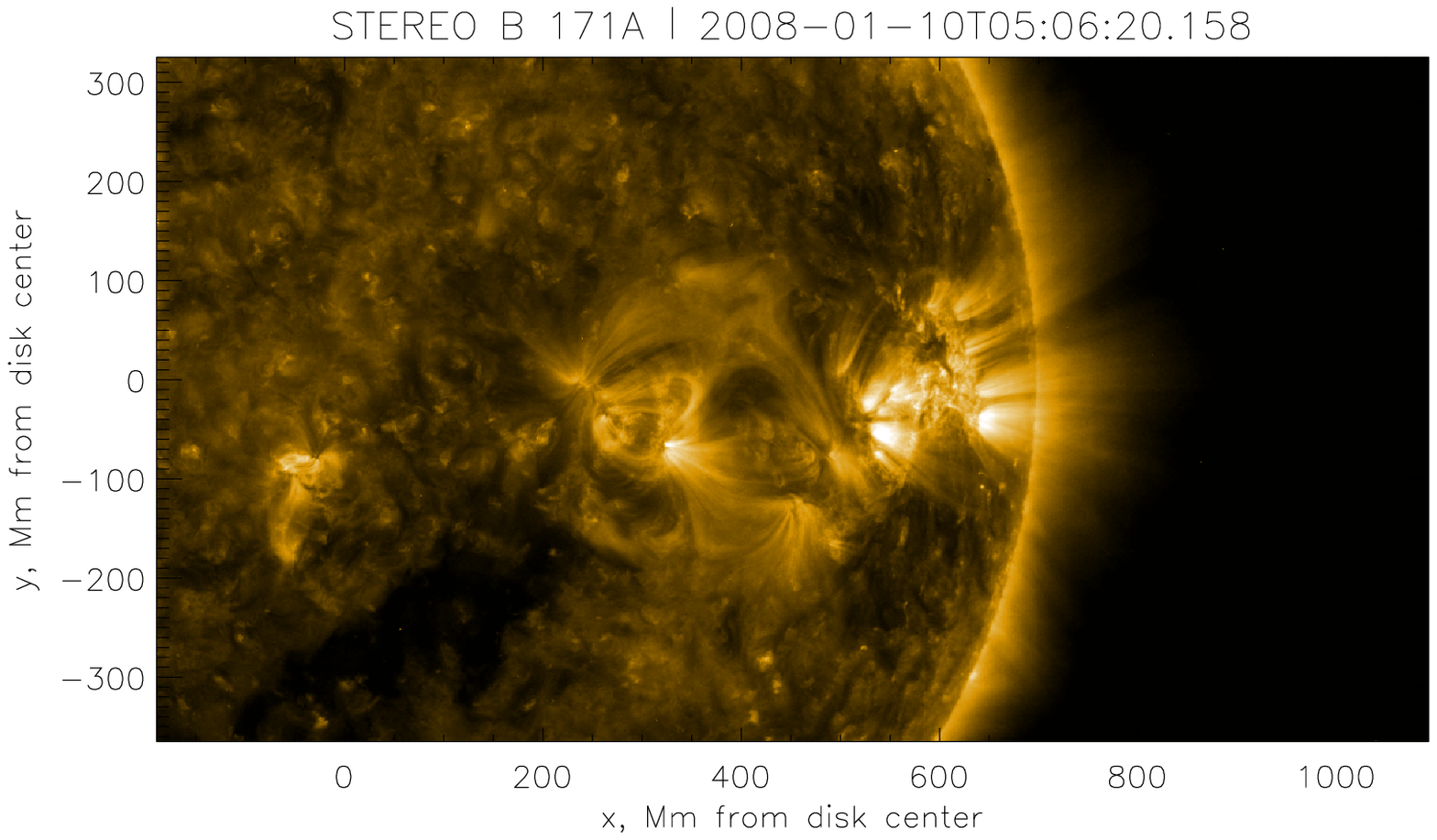} \\ 
  \end{tabular} 
 \end{center} 
 \caption{STEREO A and B images of the same region on the Sun, in the same wavelength, about 20 seconds apart. At this date, the separation between the satellites was about $45^\circ$. The images are corrected for exposure and plotted in the same color scale. The length units are shown in Mm from disk center to account for difference in the satellites' distance from the Sun. In the next two figures we examine these images in more details.}
 \label{stereo_lg}
\end{figure}

A detailed comparison of the same set of loops viewed by \textit{STEREO} satellites might clarify this topic. However, besides the abovementioned issues with optical thinness of the corona, such analysis would be greatly complicated by the choice of the separation angle between the satellites. It must be large enough for the difference in widths to be observable, yet small enough for the loops to be still identifiable on both images. Attempting such a study is not the intention of this paper; however we would still like to provide a qualitative example of such behavior of coronal structures. Fig.~\ref{stereo_lg} shows images from the two \textit{STEREO} satellites taken at nearly the same time (on 2008-01-10T05:06, less than a minute apart when corrected for the light travel time and in the same wavelength of 171\AA\refereetwo{)}. At the time these images were taken the separation angle between the satellites was about $45^\circ$. 

\begin{figure}[!hc] 
  \begin{center} 
   \begin{tabular}{cc}
     \includegraphics[height=5cm]{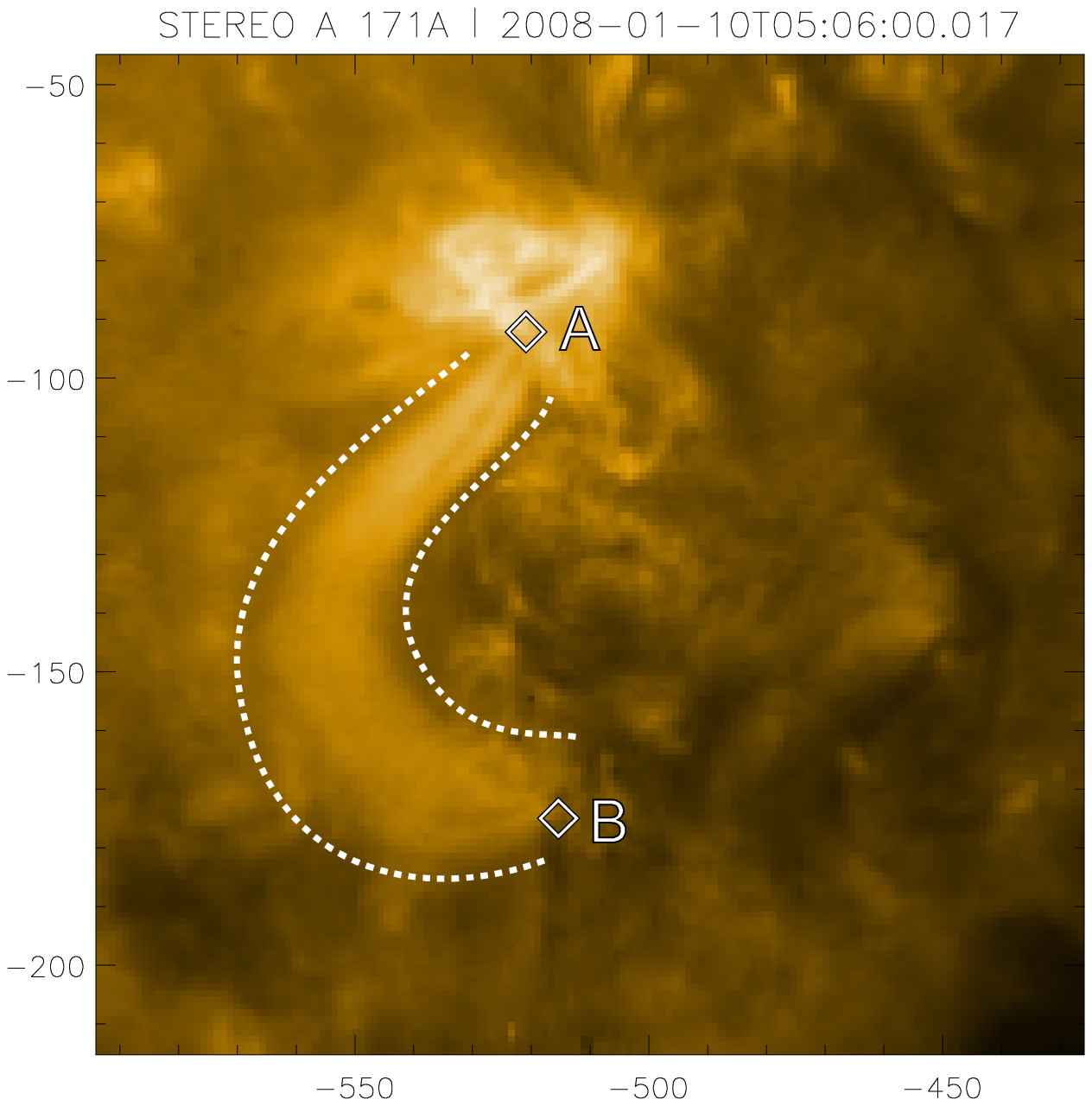} & \hspace{.5cm}\includegraphics[height=5cm]{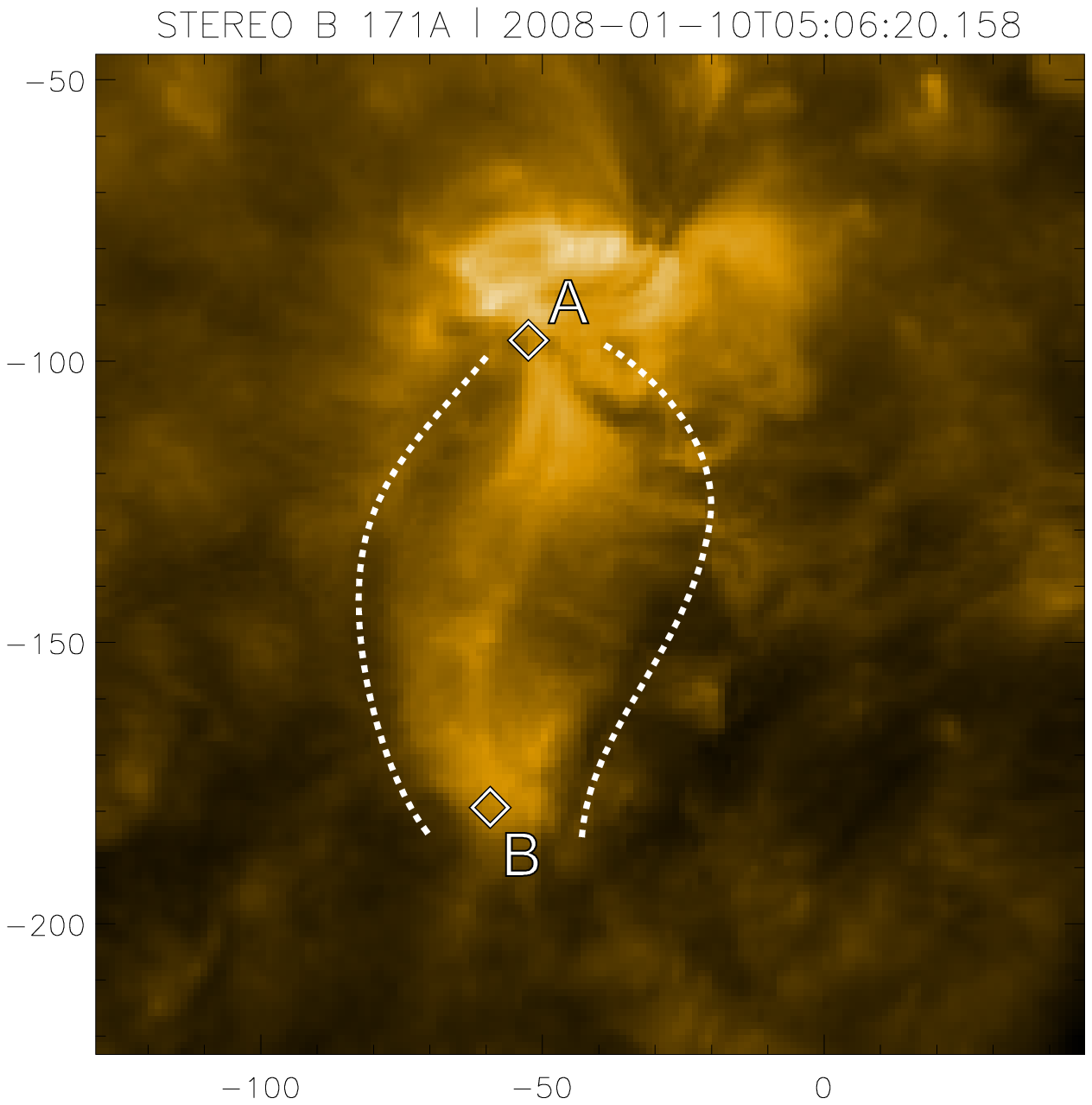} \\ 
  \end{tabular} 
 \end{center} 
 \caption{A portion of the images from Fig.~\ref{stereo_lg} showing a bundle of loops in an ephemeral active region. To guide the reader, two footpoints of the bundle, A and B, are marked on the solar surface (they were selected on STEREO A image and remapped to the STEREO B coordinates). The bundle appears to have very different shape on these two images, as outlined by the white dotted lines. If the bundle was circular in cross-section, it would have the same width and the same line of sight depth when viewed from the side (STEREO A) and from the top (STEREO B). It is possible that the background obscures the true shape of the bundle. It is, however, just as possible that the bundle is simply non-circular in cross-section. For example, such a difference in widths for different viewing angles is expected for a squished cylinder, whose cross-section is elongated along one line of sight and across the other one. This option is favorably supported by the fact that on STEREO A (left image) the bundle is narrower and brighter, while on STEREO B (right image) the bundle is wider and dimmer, even though the images are corrected for exposure and plotted on the same color scale.}
 \label{stereo_r1}
\end{figure}

The following two figures show close-ups of two regions on these frames. Fig.~\ref{stereo_r1} shows an ephemeral region in the east portion of the images and a loop bundle which connects it to the ambient field. Notice different shape of this bundle viewed from two different angles. We have marked approximate locations of the two footpoints of this bundle, A and B, to guide the reader, and outlined the bundle on both images. The points A and B were picked on one of the images and then remapped into the other one. If the bundle was circular in cross-section, it would have the same width and the same line of sight depth when viewed from the side (STEREO A) and from the top (STEREO B). It is possible that the background obscures the true shape of the bundle. It is, however, just as possible that the bundle is simply non-circular in cross-section. For example, such a difference in widths for different viewing angles is expected for a squished cylinder, whose cross-section is elongated along one line of sight and across the other one. This option is favorably supported by the fact that on STEREO A (left image) the bundle is narrower and brighter, while on STEREO B (right image) the bundle is wider and dimmer, even though the images are corrected for exposure and plotted on the same color scale. 

\begin{figure}[!hc] 
  \begin{center} 
     \includegraphics[height=7cm]{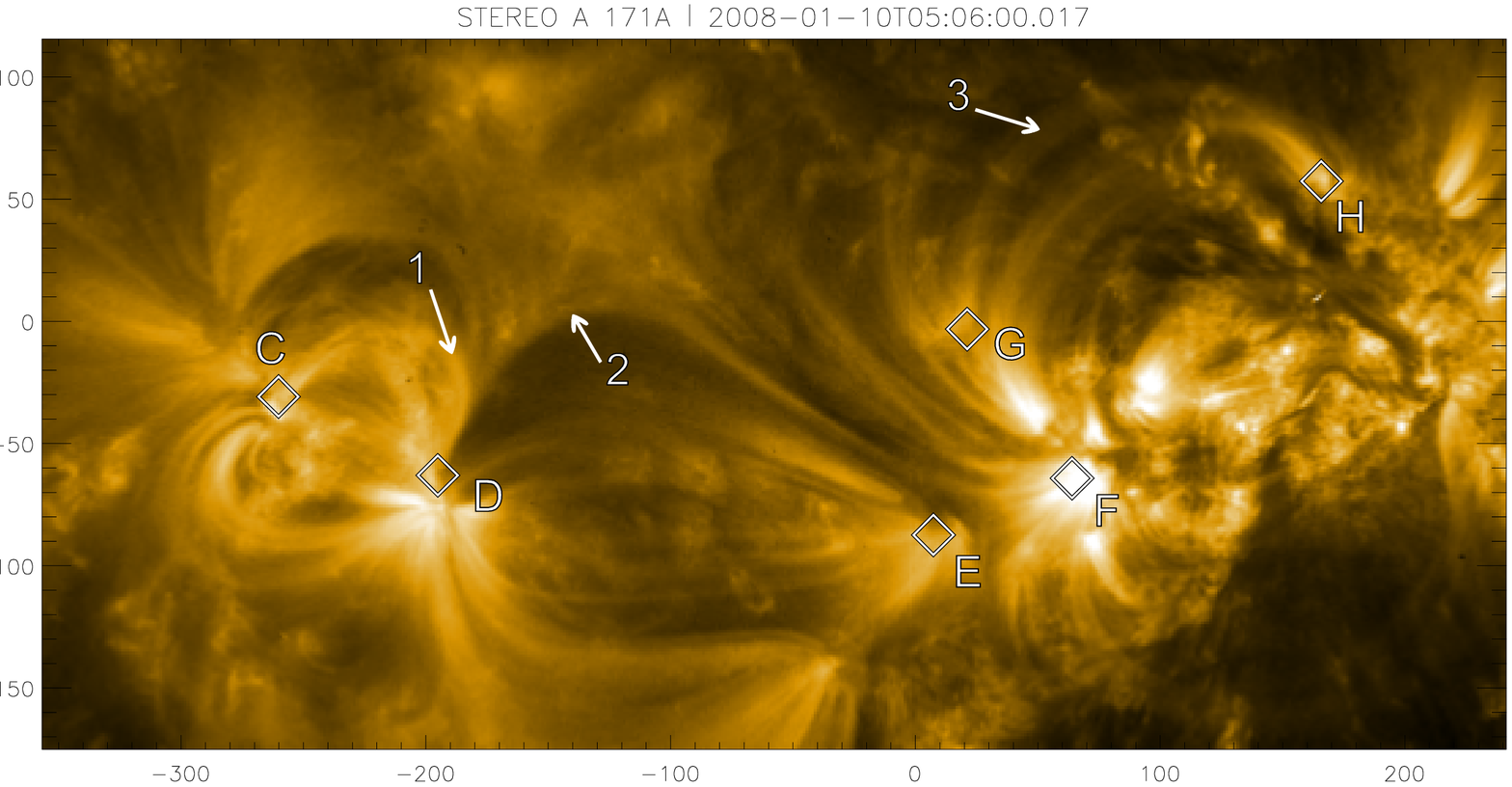} \\ \vspace{1cm} \includegraphics[height=7cm]{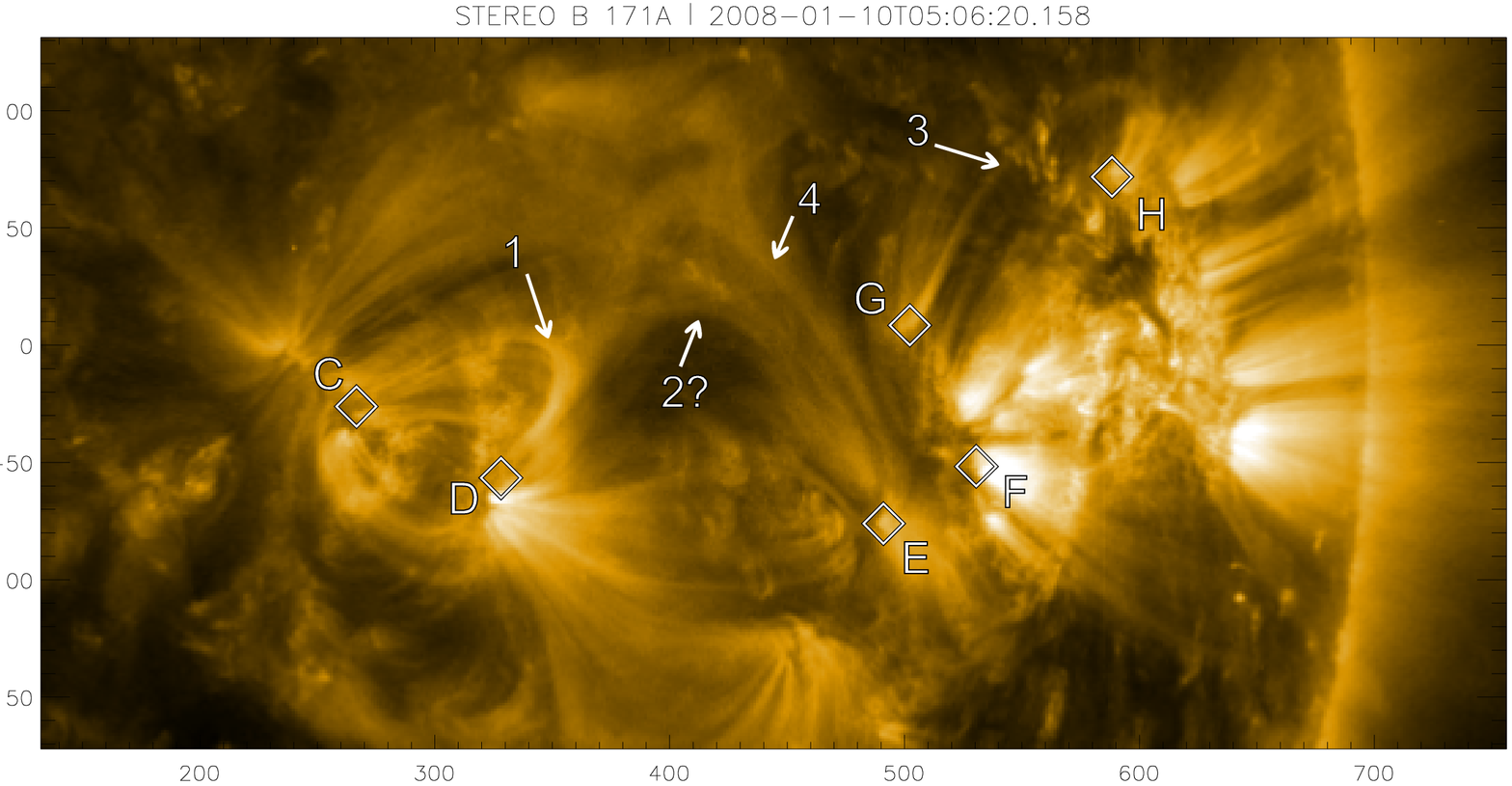}
 \end{center} 
 \caption{Another portion of the images from Fig.~\ref{stereo_lg}. As before, several points, C-H, are marked on the solar surface to guide the reader. The arrows and the numbers 1-4 point at several coronal features. Loop 1 appears much thinner in the apex on STEREO A (left panel) than on STEREO B (right panel). Loop 2, which clearly connects points D and F on STEREO A, is nowhere to be found on STEREO B (a possibly corresponding feature is shown with a question mark; however, its footpoints are clearly D and E, not D and F). On the contrary, loop 4, connecting points D and E and extending more northwards than point G on STEREO B, is hard to find (if at all possible) on STEREO A. Loop 3, connecting points G and H, appears much thinner on STEREO B than on STEREO A. Again, while it could be argued that all of these could be explained by overlap of many features in optically thin plasma, it is also possible that these loops are not circular in cross-section.}
 \label{stereo_r2}
\end{figure}

Fig.~\ref{stereo_r2} shows two active regions and their interconnecting loops. As on the previous image, several points, C-H, were marked at the surface of the Sun to guide the reader. The arrows and the numbers 1-4 point at several coronal features. Loop 1 appears much thinner in the apex on \textit{STEREO} A (left panel) than on \textit{STEREO} B (right panel). Loop 2, which clearly connects points D and F on \textit{STEREO} A, is nowhere to be found on \textit{STEREO} B\refereetwo{. A possibly corresponding feature is shown with a question mark; however, its footpoints are clearly D and E, not D and F.} On the contrary, loop 4, connecting points D and E and extending more northwards than point G on \textit{STEREO} B, is hard to find (if at all possible) on \textit{STEREO} A. Loop 3, connecting points G and H, appears much thinner on \textit{STEREO} B than on \textit{STEREO} A. Again, while it could be argued that all of these could be explained by overlap of many features in optically thin plasma, it is also possible that these loops are not circular in cross-section.

In this study, we revisit the topic of the cross-sectional shape of coronal loops. We demonstrate that even flux tubes that are approximately round at some location, do not maintain this shape further along their length in the corona. We also show how unawareness of this fact might affect our interpretation of coronal observations. In Section~\ref{sec_math} we argue that there are no mathematical reasons for flux tubes to maintain the same cross-sectional shape --- much in agreement with these arguments of \citet{Gudiksen2005b}, but in more detail. In Section~\ref{sec_anisot_stats} we quantify the anisotropy in expansion of flux tubes, using a model field based on an HMI magnetogram. Section~\ref{sec_2loops} demonstrates that \refereetwo{the amount of oblateness can be} sufficient to prevent us from observing expansion of loops and that selection criteria make the loops which lack apparent expansion appear brighter on coronal images. In Section~\ref{sec_scale_height} we show that anisotropy of expansion might be responsible for higher \textit{apparent} pressure scale height even in resolved coronal loops. Finally, Section~\ref{sec_discussion} summarizes our findings and discusses their potential implication for our understanding of the structure of the corona and the mechanisms that heat it. 

\clearpage
\section{Squashing factor --- in the corona}\label{sec_math}

The mapping, which magnetic field lines provide between two surfaces normal to the field, is not in general shape-conserving. This is usually described through a squashing factor $Q$, which is a dimensionless number normalized to be unity for shape-conserving mapping and getting bigger with increasing distortions \citep{Titov2002b}. 

The squashing factor $Q$ is usually used to describe the mapping between regions of flux of positive and negative polarity \textit{on the photosphere}. Its value is infrequently used in studies of magnetic topology, as its qualitative behavior sufficiently marks the location of topological features: $Q\rightarrow\infty$ defines separators and $Q\gg 1$ defines quasi-separator layers or QSLs \citep{Titov2002}. 

\begin{figure}[!hc] 
  \begin{center} 
     \includegraphics[width=12cm]{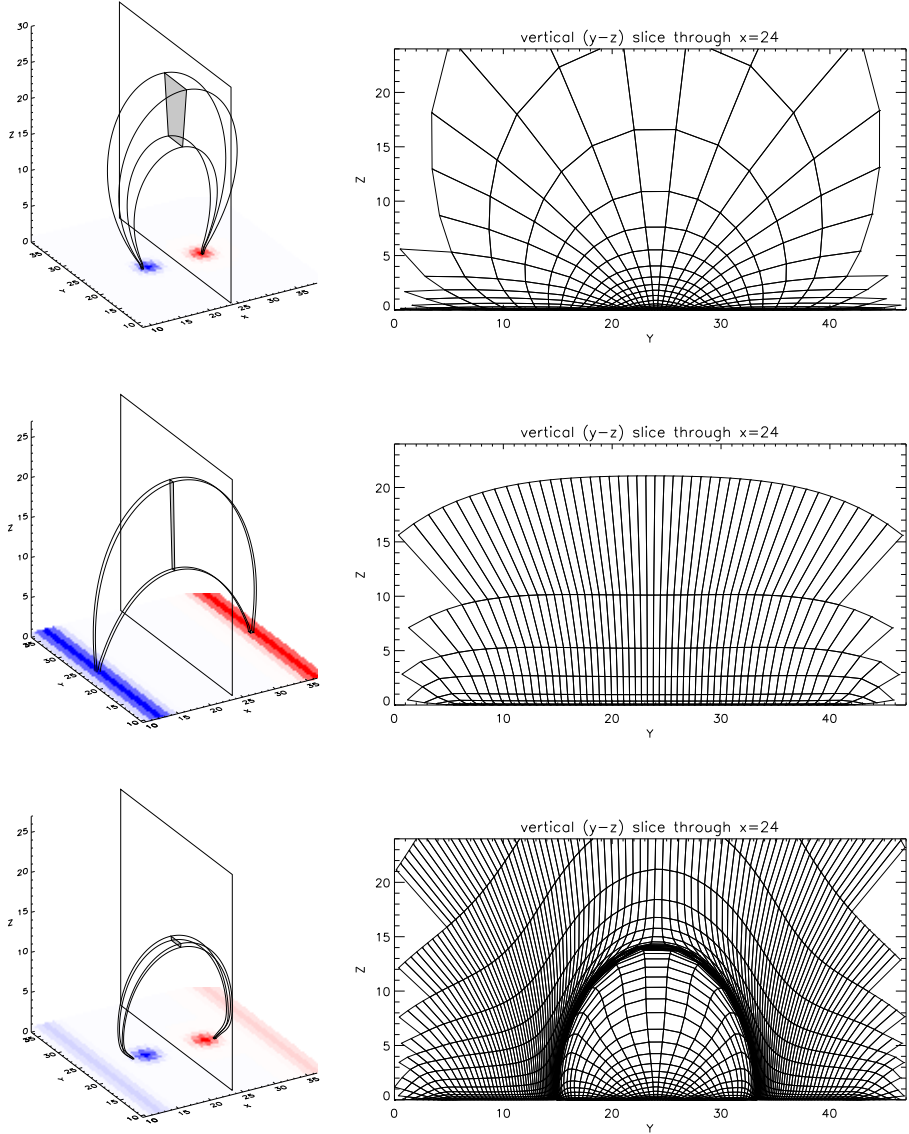} 
  \end{center} 
 \caption{Mapping of field lines from the lower boundary to the vertical plane for several simple potential fields: a dipole, an arcade and a dipole between the poles of an arcade. In each case, field lines were initiated at the lower boundary from the positive polarity, in the corners of a \textit{square} grid (with spacing less than one pixel on the lower boundary). On the left, we plot four such field lines from the corners of one square at the lower boundary. If we define a flux tube to have square cross-section at the lower boundary with these field lines in the corners of the square, the cross-section of this flux tube in the apex would correspond to the shaded shape on the middle plane, which is simply a rectangle with corners set by the intersection of the ``corner'' field lines with this plane. Note how different are the cross-sections for three different cases -- and how different are they all from the starting square at the lower boundary! The right column shows the mapping of many field lines initiated from square grid onto this middle plane (the nodes on this plot are where the field lines intersect this plane and the adjacent nodes are connected showing the cross-sections of flux tubes with square base at the lower boundary).}
 \label{3d_simple_cases}
\end{figure}

Defined this way, the squashing factor is not, in general, informative for the cross-sectional distortions of flux tubes \textit{in the corona}. We illustrate this on several simple examples shown on Fig.\ref{3d_simple_cases}. These are three potential fields confined to half space with the lower boundary being a dipole (top row), an arcade (middle row) and a superposition of the two (bottom row). Due to the symmetry of these fields, the squashing factor for the photosphere-photosphere mapping is unity for all three of them (that is, a square on one polarity maps to a square on the other polarity). The left column shows four field lines initiated from the corners of a small square on the photosphere and the shape they map to at the apex. The right column shows the mapping set by the magnetic field from the lower boundary to the vertical plane in the middle of the computational domain (again, due to symmetry of the system the field is normal to this plane at all points). We initiated a set of field lines from the corners of a \textit{square} grid at one polarity at the lower boundary and found a set of points where these field lines cross the vertical plane. Adjacent points are connected by straight lines which in turn form a grid that the square grid at the photosphere maps to. The grid has gaps on the edges as we did not plot the grid cells which were partially outside of the domain. The top boundary on these plots is slightly below the boundary of the domain. Note that many grid cells on the vertical plane are substantially deformed. The corresponding flux tubes, while having a square base, have elongated cross-section at the apex. Note also that properties of the deformation depend on the configuration of the sources in the lower boundary. For example, in the arcade, flux tubes \textit{cannot} expand with height along the translational direction of the arcade, so the expansion must take place entirely in the vertical direction (the arcade in our example is finite so this effect vanishes towards its edges). 

\begin{figure}[!hc] 
  \begin{center} 
     \includegraphics[width=16cm]{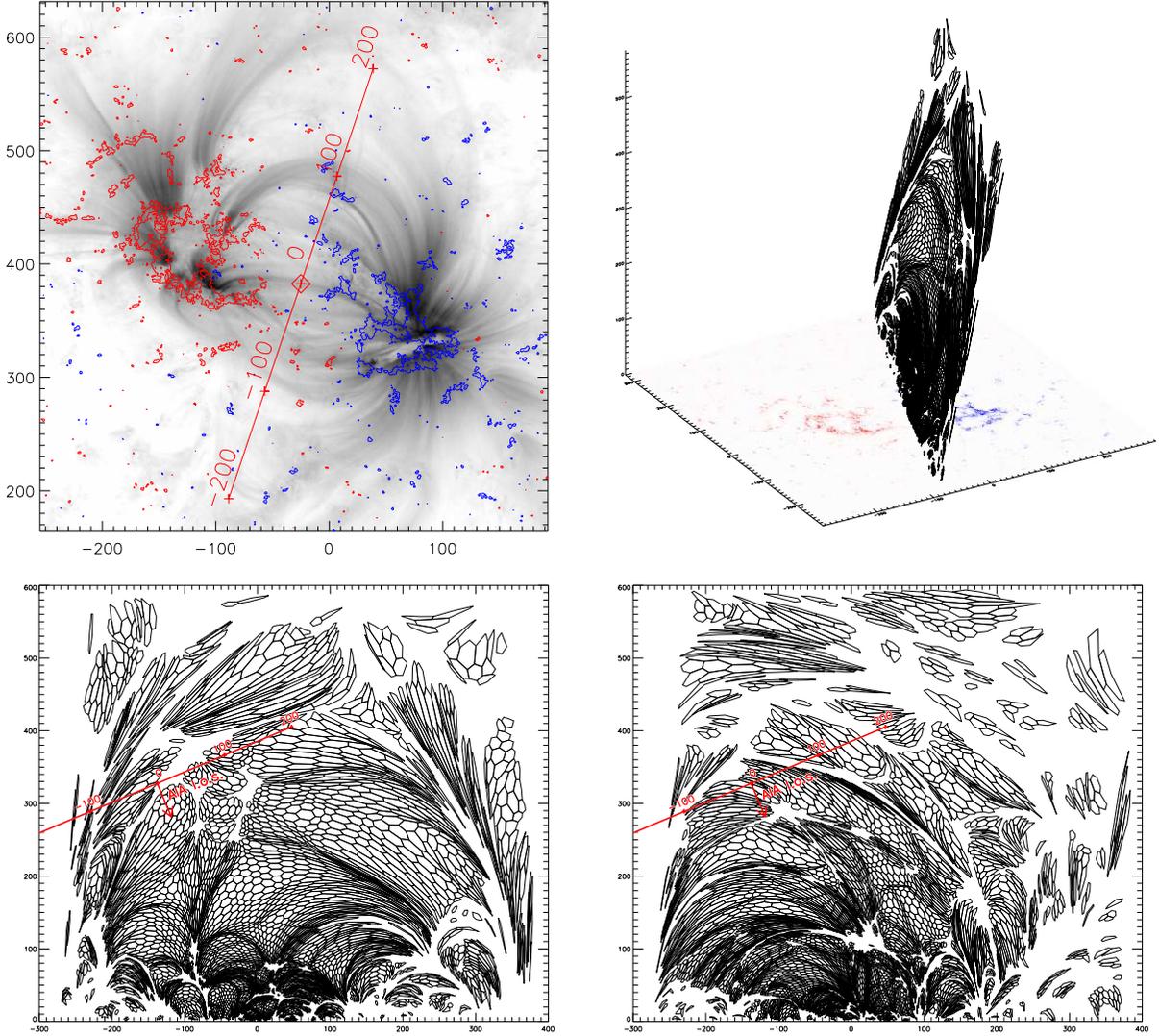} 
  \end{center} 
 \caption{Mapping of field lines from the lower boundary to a vertical plane in a manner similar to Fig.~\ref{3d_simple_cases}. In this case, field lines were initiated from corners of regular hexagons at the lower boundary. The plane was set by AIA line of sight and an arbitrary line in the plane of the sky which visually passes through tops of many loops (the red line on the top left panel, showing AIA 171A image and the corresponding HMI magnetogram). The top right panel shows the position of this plane in the 3D computational domain. The points at which the field lines cross this plane are calculated and plotted for the field lines initiated from negative (bottom left) and positive (bottom right) polarities at the lower boundary. Adjacent points are connected to form polygons --- these would correspond to cross-sections of flux tubes in the active region, should these flux tubes be regular hexagons at the lower boundary. The gaps in the plots reflect that we did not trace field lines from where $|Bz|$ was below a 100G threshold. We encourage readers of the electronic edition to zoom in the two bottom panels to examine cross-sections of the low-lying flux tubes. Very few flux tubes stay could still be described as regular hexagons at the apices (here we mean more of a qualitative description of a visual impression, and we make quantitative estimates in Section~\ref{sec_anisot_stats}). As we argue in Section~\ref{sec_2loops}, it may have substantial impact on the visibility of the corresponding coronal loops.}
 \label{2010-08-15_slice}
\end{figure}

These three examples correspond to very simple fields. For more complex boundaries, the distortions of the cross-section would in general be even stronger. Fig.~\ref{2010-08-15_slice} shows a similar mapping for a potential field based on HMI magnetogram onto a plane parallel to the line of sight. \referee{We analyzed AR NOAA 11097 on 2010-08-15. (We have no reasons to consider it a special active region in any sense. The choice was based on the visibility of loops, proximity to central meridian and availability of high-resolution SDO/HMI data, as we further argue on the importance of the fine-scale structure.)} The top left panel shows where this plane crosses plane of sky and top right panel shows where is it located in the computational domain (to construct the field, the magnetogram is remapped to disk center and downsampled by a factor of 2). Much like in the previous example, we initiate field lines at the corners of a regular grid (in this case, hexagonal) and calculate where these field lines cross the given plane. Bottom left and right panels show the mapping from the positive and negative polarities, respectively. The gaps in the grid are due to that we do not initiate field lines from the points where field strength is $|B_z|<100$G. 

Fig.~\ref{2010-08-15_slice} thereby shows what the cross-section of flux tubes look like along the line of sight, if their bases are round. Most of them undergo substantial distortions as they expand. If coronal loops represent emission of plasma in individual flux tubes, their cross-sections must in general be \textit{anything} but circular --- even in this simple example. This effect is possibly even stronger in the actual coronal field. First, because the photosphere has a lot of fine structure that we missed by using a downsampled HMI magnetogram, and as Figs.~\ref{3d_simple_cases}-\ref{2010-08-15_slice} show, flux tubes tend to be more oblate near the boundaries of individual domains of connectivity. Second, the actual corona has currents flowing in it which might also affect the magnitude of distortion --- not necessarily decreasing it. For example, current sheets are associated with regions of increased squashing \citep{Buchner2006}, so flux tubes in non-current-free corona might in fact be even more oblate than in the example that we show. But even these amounts, as we discuss further in the text, suffice to cause substantial impact on our interpretation of coronal observations.

Here and further in the text we make the assumption that the base is round, but this is done merely for simplicity of the argument. Indeed, suppose the flux tubes are round at the apex instead. The mapping from the apex to the base is described by the inverse of the one from base to the apex and so the footpoints of flux tubes would have to be oblate. Likewise, a flux tube with round cross-section at a given point along its length, would in general be oblate at other points.

\clearpage
\section{Statistics}\label{sec_anisot_stats}

We further investigate how much flux tubes deform at the apex if their cross-section at the base is circular. To do this, we perform the following exercise. We consider a set of flux tubes that fill the entire cube of the field from Fig.~\ref{2010-08-15_slice} (less the volume with field strength at the base below 100G threshold) in the manner similar to that on Figs.~\ref{3d_simple_cases}-\ref{2010-08-15_slice}. We initiate field lines on the positive (or the negative) polarity at the nodes of hexagonal grid. The hexagons they define are considered bases of flux tubes. We further follow these flux tubes along their axes and examine what is their shape at the point of maximal expansion. The shapes on Fig.~\ref{2010-08-15_slice} are cross-sections of flux tubes, but not necessarily perpendicular to the axis, neither in general are they located at the point of maximal expansion, so for a more rigorous approach we develop a different calculation.  

\begin{figure}[!hc] 
  \begin{center} 
   \begin{tabular}{cc}
    \includegraphics[width=7cm]{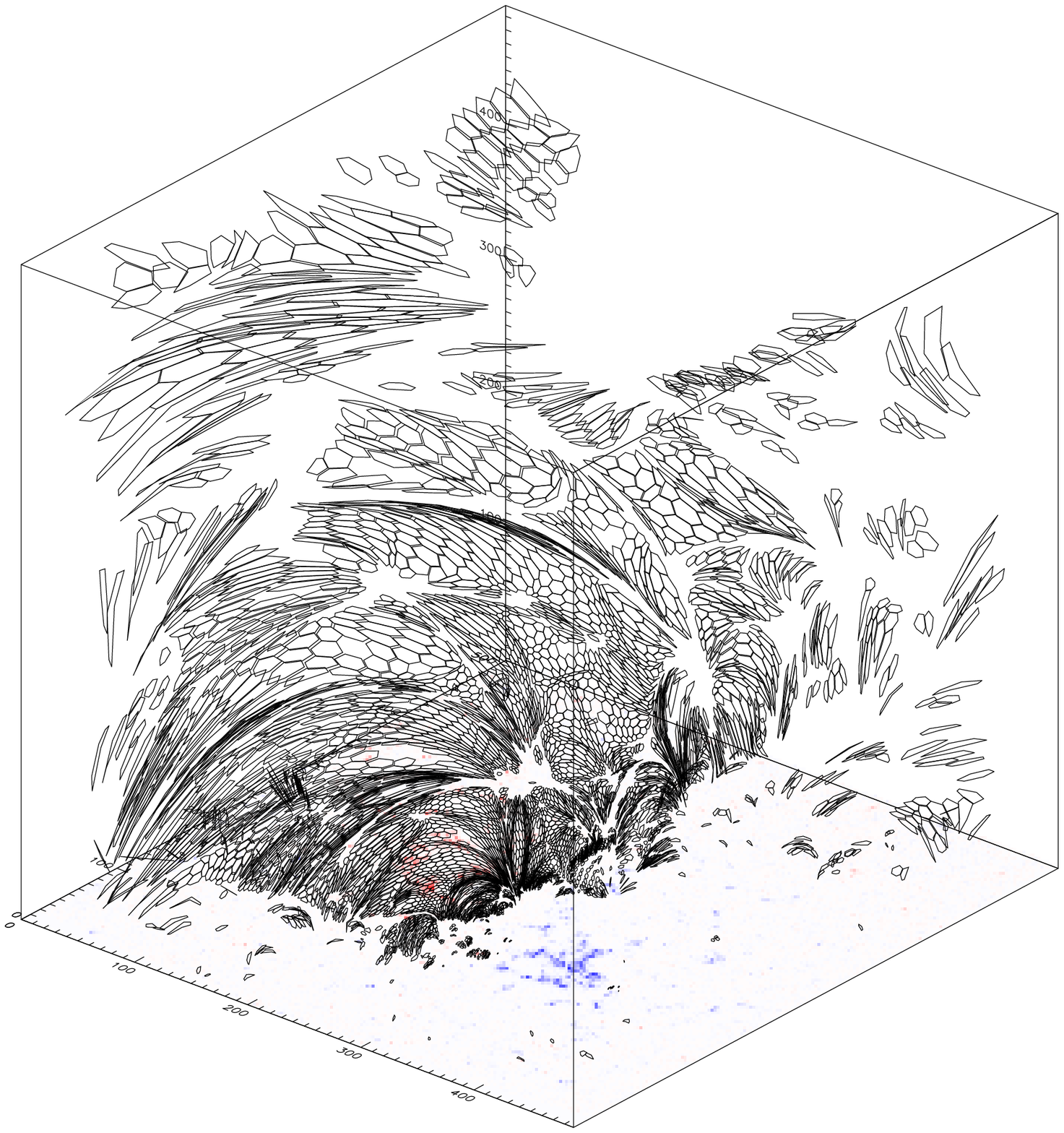} & \hspace{0.5cm}\includegraphics[width=7cm]{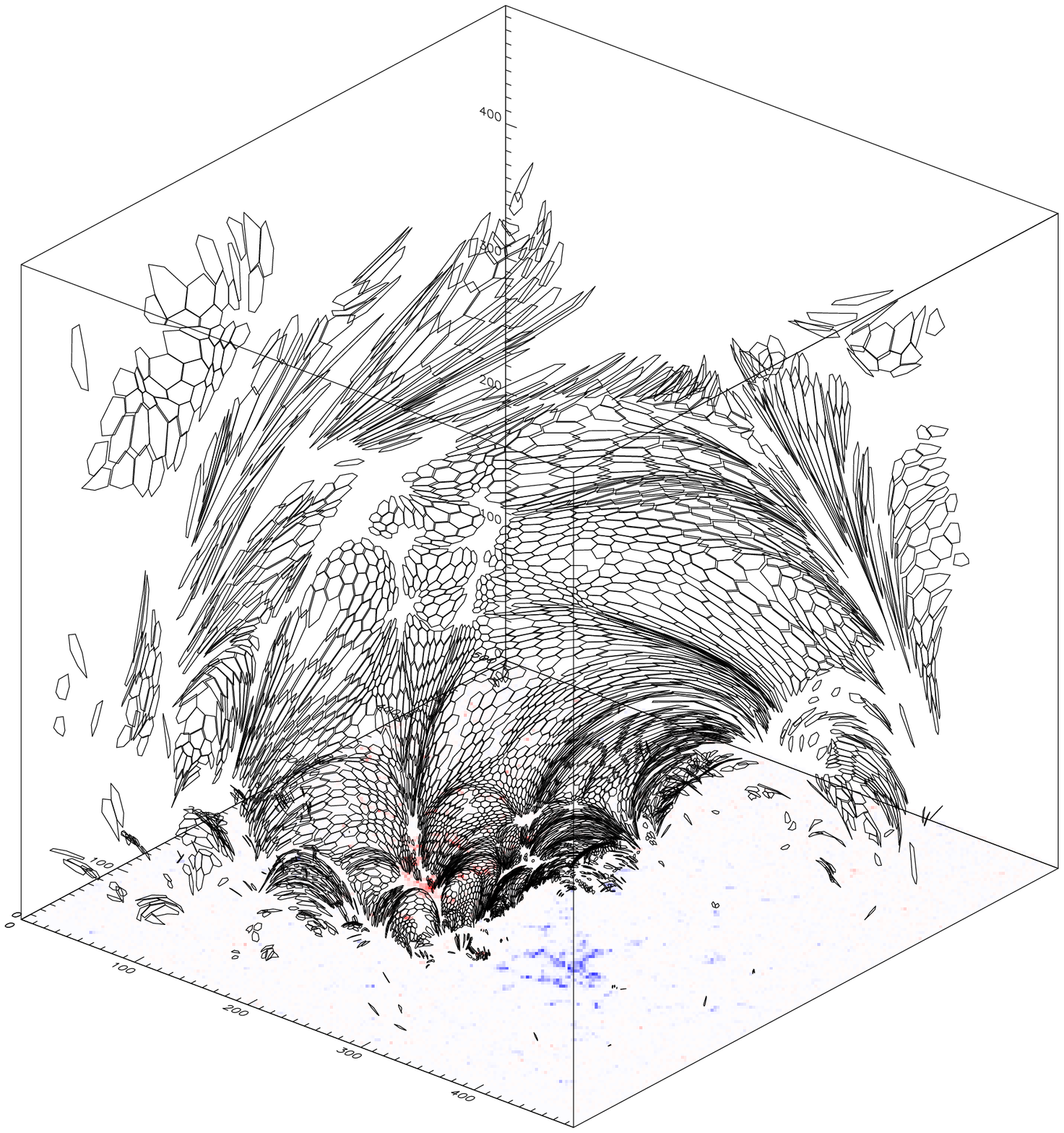} \\ 
  \end{tabular} 
 \end{center} 
 \caption{The same set of magnetic field lines as on Fig.~\ref{2010-08-15_slice}, but this time every individual set of six field lines was cut at the apex by a plane perpendicular to the direction of the field at the central field line. We define apex as the point along a flux tube where the field strength is minimal and therefore cross-sectional area of the flux tube is maximal. (These plots may  be viewed from different angles in the animations available in online materials.)}
 \label{apex_mesh}
\end{figure}

We perform further analysis as follows. For each flux tube, we initiate an axis field line at the center of the hexagonal base and define the ``apex'' point as the point of maximal expansion (it is calculated as the point with minimal field strength along the axis). We then construct a plane perpendicular to the axis at this point and calculate where the six corner field lines cross this plane. The result is a planar hexagonal polygon (not in general regular). These shapes are shown on Fig.~\ref{apex_mesh} for flux tubes starting from positive (left panel) and negative (right panel) polarities. Online materials include the animation of these plots being rotated to show them from different sides. As before, the gaps in the plots correspond to the areas for which field strength at the base was below a 100G threshold. 

\begin{figure}[!hc] 
  \begin{center} 
    \includegraphics[width=8cm]{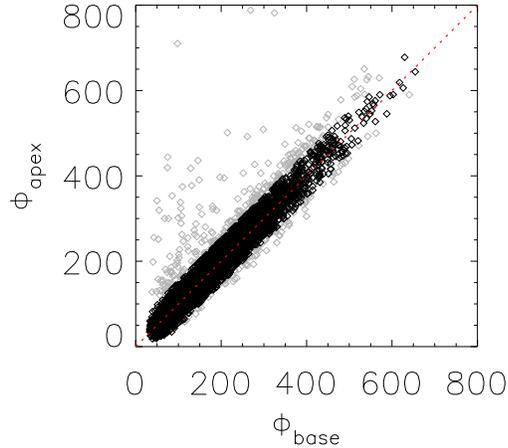} 
 \end{center} 
 \caption{Comparison of flux at the bases and at the apex of the six-sided flux tubes. It demonstrates that flux is well conserved for most of the ``tubes'' and therefore the approximation we make for their cross-sectional shape holds for most of them. For further analysis we discard those for which the fluxes differ by more than 50~$\mbox{G}\cdot\mbox{pix}^2$ (8636 flux tubes out of the total of 9050 or about 95\%). The 414 discarded points are plotted in gray and the remaining are plotted in black.}
 \label{apex_fluxconserv}
\end{figure}

The estimated magnetic flux is well conserved along these model flux tubes, as demonstrated on Fig.~\ref{apex_fluxconserv}. It shows a scatter plot of magnetic flux at the apex of each flux tube versus that at the base. In both cases we use the thin-tube approximation and assume that the field strength does not vary along the cross-section and the cross-sectional area to be approximately that of the polygon with corners at the six corner field lines. Possible reasons for flux variation along the flux tube could then be, for example, substantial variation of magnetic field strength on a cross-section, inaccuracies in numerical integration of field lines (especially around topological divides) and the shape of the cross-section becoming too distorted to be adequately treated as a polygon. We find that these effects are small for the majority of flux tubes, as shown on Fig.~\ref{apex_fluxconserv}. We further concentrate on 8636 flux tubes out of 9050 (about 95\%) for which the discrepancy between flux at the apex and flux at the base is less than 50G$\cdot\mbox{pix}^2$. 

The cross-sections of flux tubes at the apices are then fit with an ellipse using a least-squares method \citep{ellipse_fit}, yielding two semiaxes $a_{apx}$ and $b_{apx}$ (where $a_{apx}>b_{apx}$). We also calculate what the width of each flux tube would have been if the expansion were isotropic as $r_{apx}=\sqrt{A/\pi}$, where $A$ is the cross-sectional area. 

\begin{figure}[!hc] 
  \begin{center} 
   \includegraphics[height=5cm]{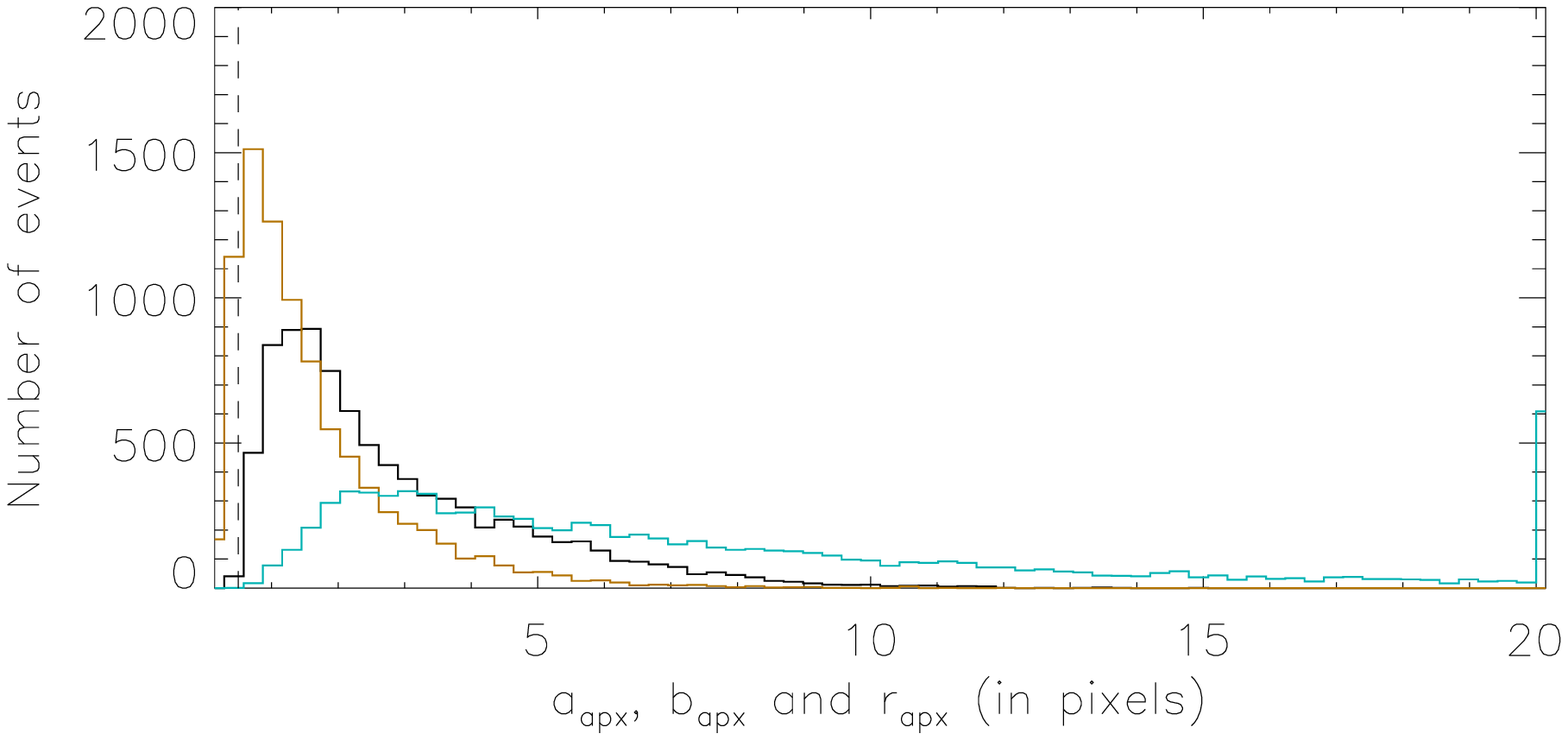} \\      
   \includegraphics[height=5cm]{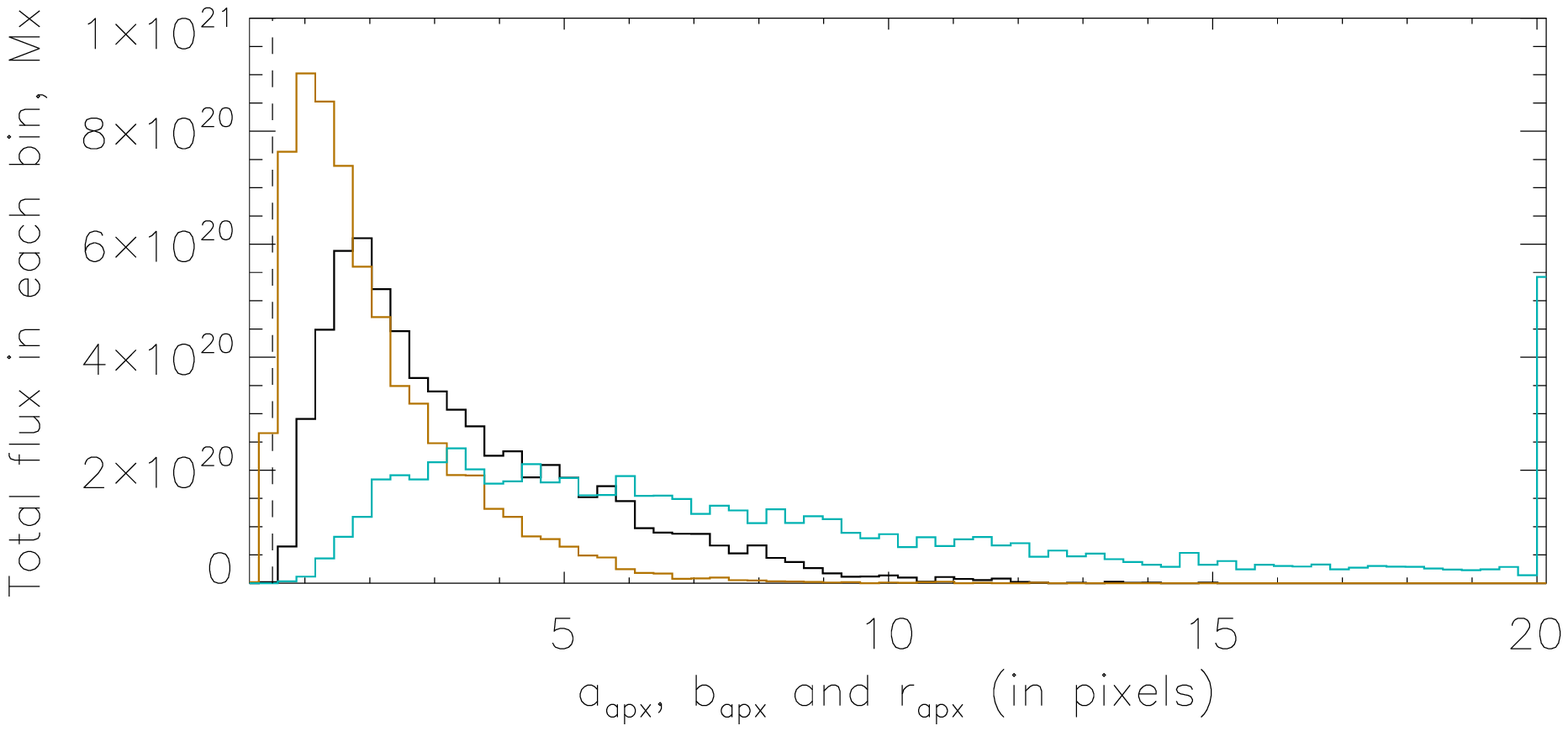} \\   
  \end{center}
  \caption{Statistics on apex expansion in elongated (cyan) and shortened (orange) directions. The top plot shows number of flux tubes which fall in the $a_{apx}$ and $b_{apx}$ bins and the bottom plot shows total flux in each bin. The black curve for each plot corresponds to what the expansion would have been if cross-section was circular, that is, it corresponds to $\Gamma=\sqrt{A_{apx}/A_0}$. The cross-sectional radius at the base is $r_0=a_0=b_0=0.5$ pix and is shown with the dashed vertical line.}
  \label{apex_stats_abg}
 \end{figure}

\begin{figure}[!hc] 
  \begin{center} 
   \includegraphics[height=5cm]{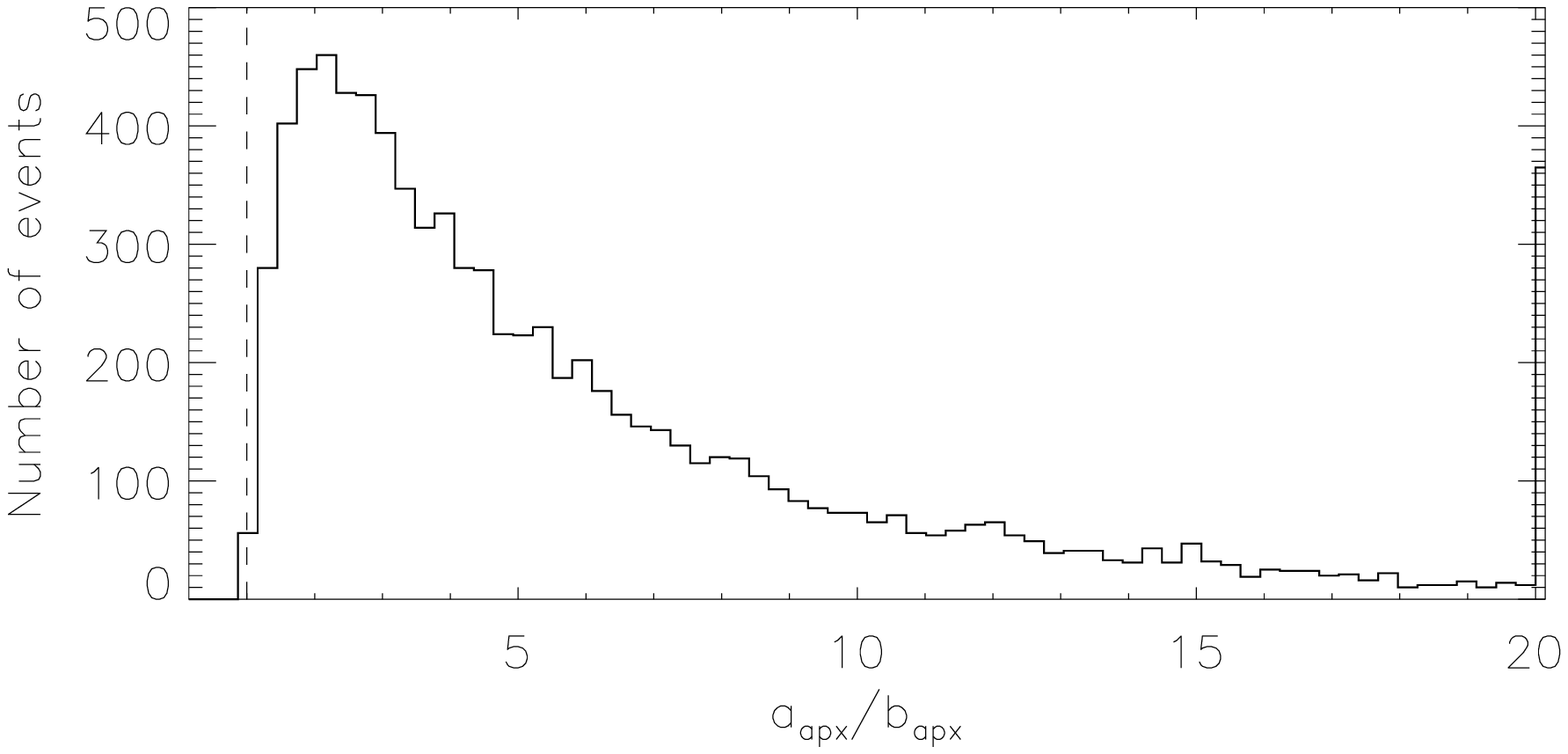} \\      
   \includegraphics[height=5cm]{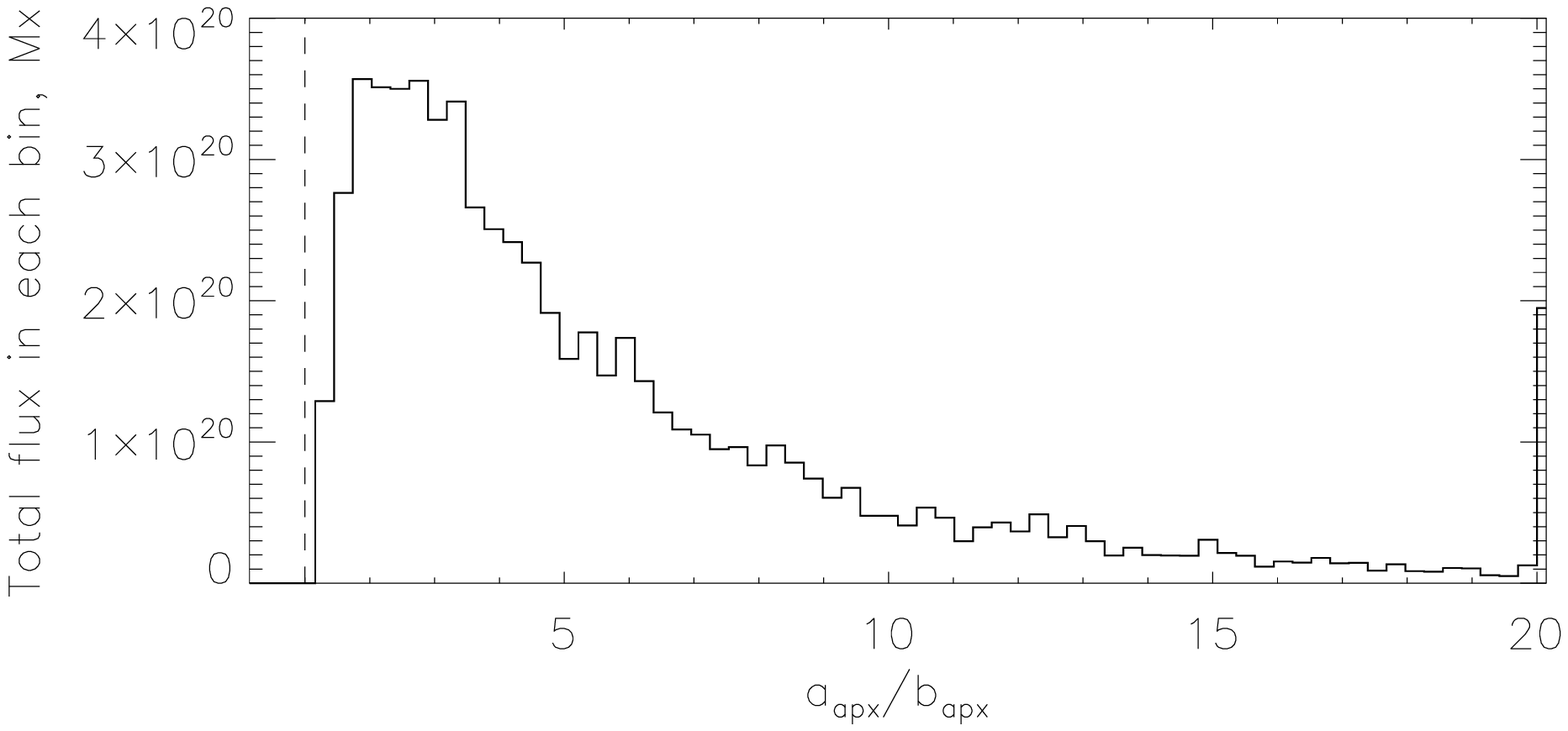} \\   
  \end{center}
  \caption{Statistics on the shape of the flux tubes at their apices. The top plot shows number of flux tubes which fall in the $a_{apx}/b_{apx}$ bins and the bottom plot shows total flux in each bin. Dashed vertical lines mark unity.}
  \label{apex_stats_ab}
 \end{figure}

The results are summarized on Figs.~\ref{apex_stats_abg}-\ref{apex_stats_ab} and Table~\ref{table_stats}. Fig.~\ref{apex_stats_abg} shows histograms of $a_{apx}$ (cyan), $b_{apx}$ (orange) and $r_{apx}$ (black). The top panel shows a histogram in the conventional sense, that is, it is a number of flux tubes (all of which have the same base areas) in each bin and the bottom panel shows total flux in all flux tubes in the same bin. The semiaxes and \refereetwo{radii} are given in pixels, and the base of each flux tube is a hexagon of $r_0=a_0=b_0=0.5$pix. On both plots it is clear that flux-tube expansion is substantially different in two directions normal to the field: a typical (in full width at half max sense) expansion factor is about 1-4 times in $a_{apx}$ and about 3-20 in $b_{apx}$, while if oblateness would unaccounted for, the typical expansion factors would have been around 2-5 times. Fig.~\ref{apex_stats_ab} shows the histogram (top panel) and the flux distribution (bottom panel) of $a_{apx}/b_{apx}$; the typical range of this quantity is 1.5 to 5 and the median is 4.5. 

These numbers likely \textit{underestimate} the actual oblateness in the cross-section of coronal flux tubes. \referee{There are several reasons for that, some of which we list below.} One is that the actual magnetic field at the photosphere probably has finer structure than what appears on our lower boundary data (which was an HMI magnetogram downsampled by a factor of 2). By doing that we possibly missed many small-scale topological divides. As Fig.~\ref{2010-08-15_slice} shows, flux tubes become more distorted near topological divides (which is in agreement with the definition of these divides via squashing factor). The other factor contributing to a possible underestimation is that we chose potential (or current-free) field model. Current sheets are associated with additional structure in the connectivity \citep{Buchner2006} so the topological structure of the actual corona \referee{is in general} more complex than that of the potential field.

 \begin{table}[!hc]
  \begin{center}
  \begin{tabular}{cccc|ccc}
  & \multicolumn{3}{c}{Number of Flux Tubes} & \multicolumn{3}{c}{Total Flux} \\
  & Median & Peak & FWHM range  & Median & Peak & FWHM Range\\ 
  \hline
   Circular $r_{apx}$, pix & 2.4 & 1.6 & 0.7 to 2.5 & 3.0 & 1.9 & 1.3 to 3.3 \\
   Ellipse $a_{apx}$, pix & 1.3 & 0.7 & 0.4 to 1.6 & 1.8 & 1.0 & 0.7 to 2.2 \\
   Ellipse $b_{apx}$, pix & 5.6 & 3.0 & 1.6 to 6.8 & 6.4 & 3.3 & 2.1 to 8.3 \\ \hline
   Aspect ratio $a_{apx}/b_{apx}$ & 4.5 & 2.2 & 1.3 to 5.4 & 4.4 & 1.9 & 1.6 to 4.8 \\

  \end{tabular}
  \caption{Statistics on the apex expansion of flux tubes. The first conclusion is that flux tubes are substantially oblate at the apices with the typical aspect ratio of 2 to 5. The second conclusion is that most of the expansion takes place along one of the directions only. Given the base radius of $r_0=a_0=b_0=0.5$ pixels, the flux tubes expand at most by a factor of 4 in one of the directions and over a factor of 15 in the other one.}
  \label{table_stats}	
  \end{center}
 \end{table}

\begin{figure}[!hc] 
  \begin{center} 
    \includegraphics[width=10cm]{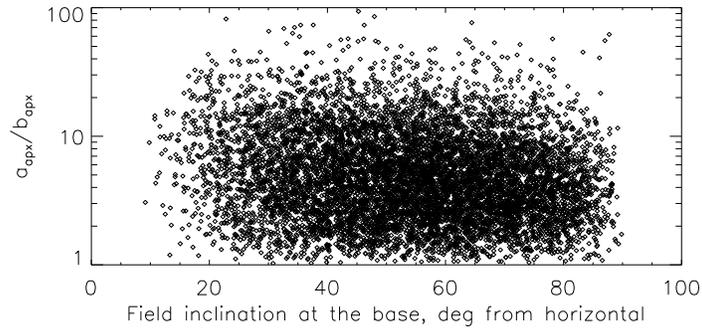} 
 \end{center} 
 \caption{The value of $a_{apx}/b_{apx}$ does not appear to be strongly correlated with the inclination of the field at the base of the flux tube. Spearman's correlation coefficient is $\rho=-0.14$.}
 \label{apex_vs_ang}
\end{figure}

Of course, the base hexagons lie in the horizontal ($z=0$) plane and so are not perpendicular to the axis of the corresponding flux tubes. It could be argued that flux tubes which are inclined at the base might in fact be oblate in the initial cross-section, thus introducing a bias in our statistics. If this bias was strong, there would be a strong correlation between the field inclination and oblateness at the apex. Fig.~\ref{apex_vs_ang} demonstrates that such a correlation is not substantial, with Spearman's correlation coefficient $\rho=-0.14$. This suggests that oblateness of a flux tube at the base does not systematically contribute to its oblateness at the apex.

\clearpage
\section{Example}\label{sec_2loops}

To better understand the effects caused by the oblateness of the loops, \referee{we simulate two coronal loops} taken from the potential field shown on Figure~\ref{2010-08-15_slice}. The bases of the flux tubes are regular hexagons of slightly different diameters so that they have similar fluxes, and they also have similar length and location within the domain. Fig.~\ref{2loops_3d} shows a 3D sketch of these two flux tubes (top left) and renderings of emission measure of two corresponding synthetic ``loops'' in three projections (the other three panels, as marked). 

\begin{figure}[!hc] 
  \begin{center}
   \begin{tabular}{m{5cm}m{5cm}} 
    \includegraphics[height=4cm]{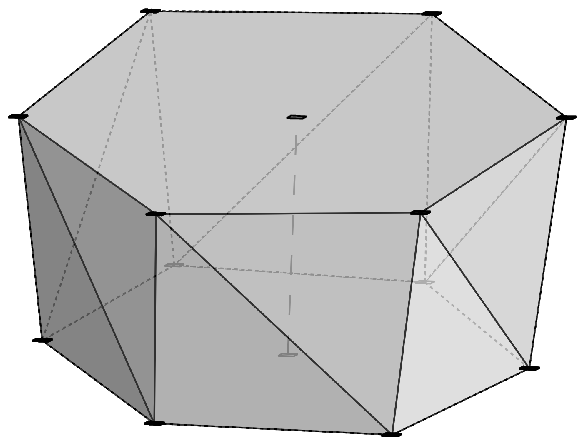}\hspace{1cm} & 
    \includegraphics[height=6cm]{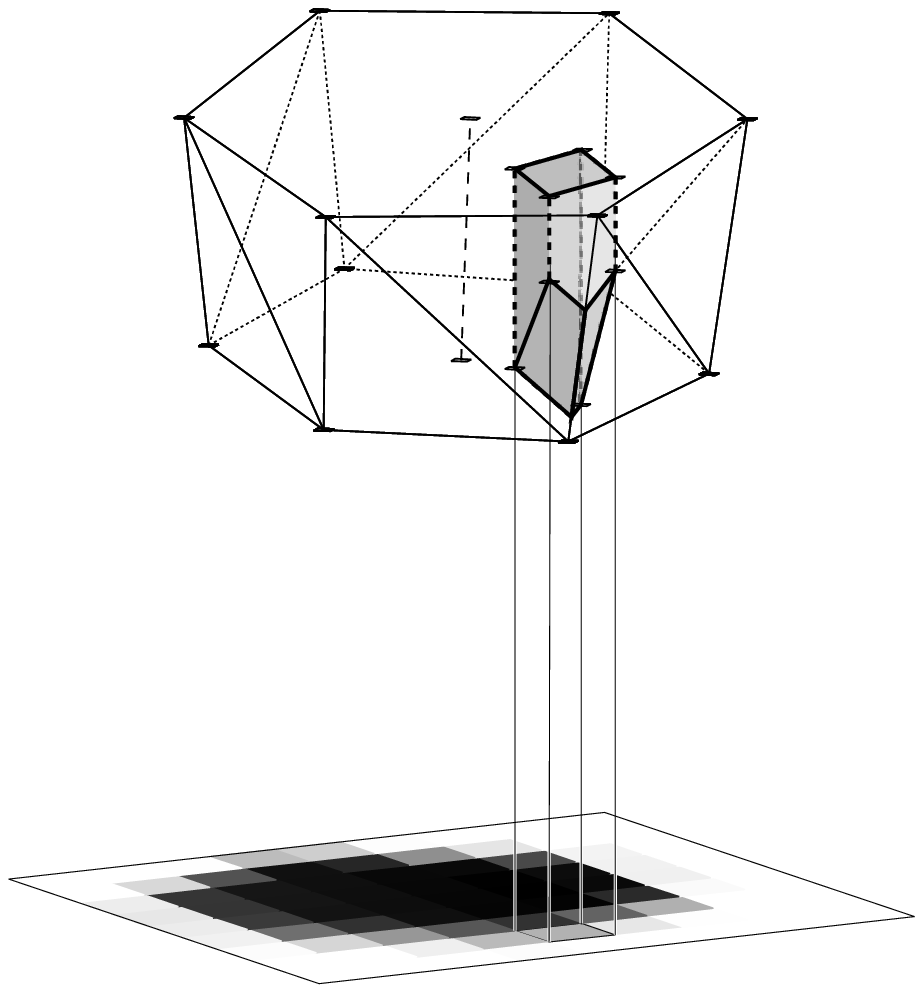} \\
   \end{tabular}
  \end{center} 
 \caption{Left panel: A schematic drawing of a ``slice'' volume along a flux tube, enclosed between two planes normal to the axis of the tube. A cross-section of a flux tube is assumed to be a hexagon (not in general regular). Field lines are traced from the corners of the base cross-section (a regular hexagon) and from the center of the base. For a given point along the axis field line (dashed line), a cross-section is set by the intersection of the plane normal to the axis with the ``corner'' field lines. Right panel: the contribution of a given slice to the rendered image intensity at a given pixel is calculated as the \textit{volume} of the portion of the slice which projects into the given pixel times the intensity multiplier of the slice (e.g., emission measure). This is an equivalent of the line-of-sight depth of the feature times its filling factor.}
 \label{fig_mesh_unit}
\end{figure}

\begin{figure}[!hc] 
  \begin{center} 
     \includegraphics[width=15cm]{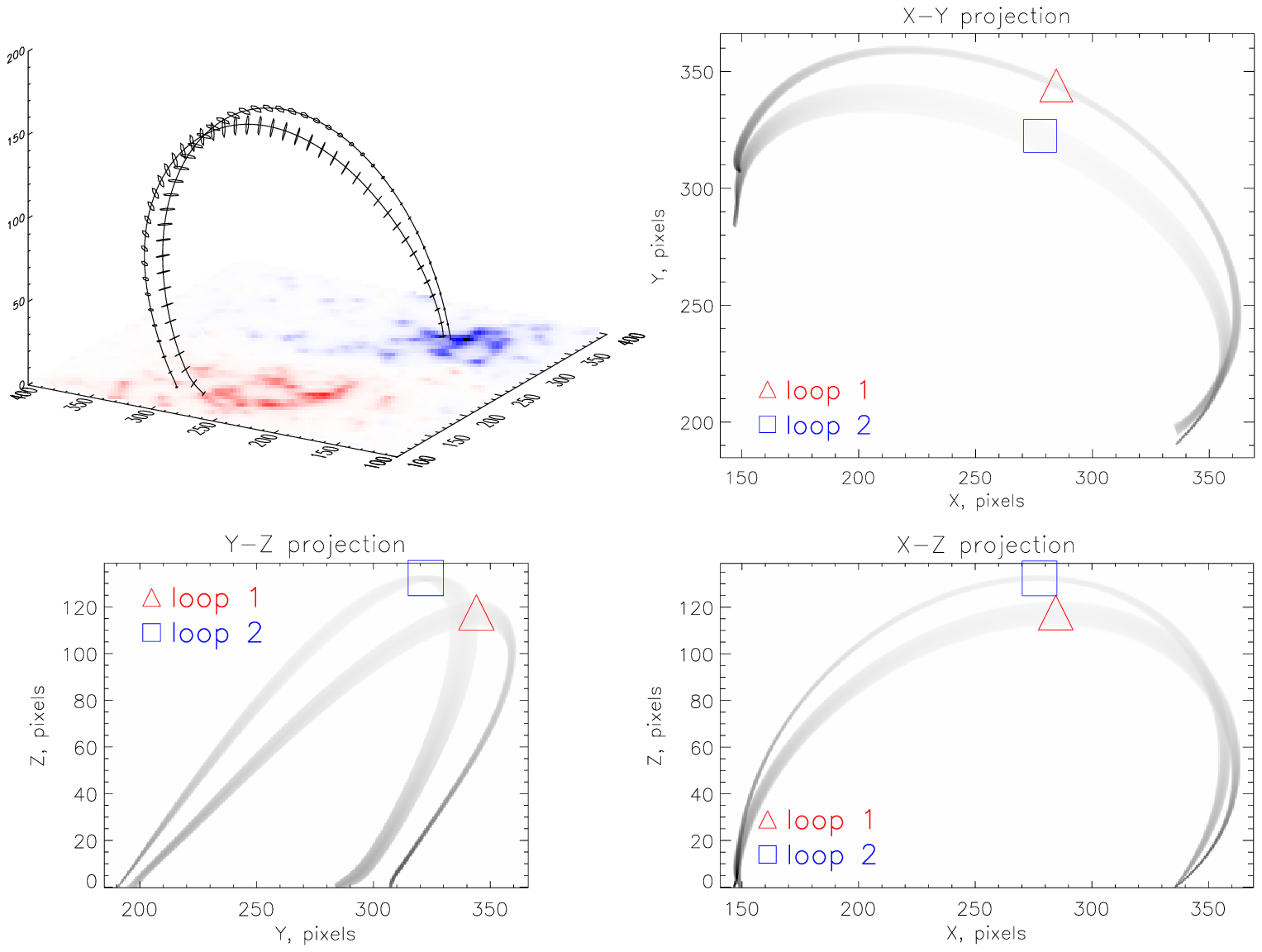} 
  \end{center} 
 \caption{A sketch of two flux tubes (top left) and the renderings of the corresponding synthetic coronal loops in $x-y$, $y-z$ and $x-z$ projections (as marked). The field lines forming these tubes were initiated at two regular hexagons at the opposite polarities. The renderings show emission measure corresponding to hydrostatic atmosphere with scale height of 40 pixels. Apex of loop 1 is marked as red triangle and apex of loop 2 is marked as blue square in all renderings. Note that loop 1 appears brighter \textit{and} non-expanding in $x--y$ projection, while the opposite happens in $x-z$ projection, see Figs.~\ref{2loops_intn}-\ref{2loops_lp2} for the numerical justification. In fact, given these two projections alone it is easy to mistakenly believe that there are two loops, one is wide and expanding and the other one is thin and non-expanding, while they in fact have similar radii and they both expand -- but in different directions. Note also how the projection effect impacts the perceived density scale height.}
 \label{2loops_3d}
\end{figure}

The flux tubes are rendered as follows. As before, we initiate six field lines from the corners of a regular hexagon and one field line in its center. For many points along this central field line we construct slices as described above and so split the flux tube into a set of thin volumes contained between consecutive slices (Fig.~\ref{fig_mesh_unit}, left panel). Each volume is assigned an ``intensity'' multiplier (e.g., emission measure). The contribution of this volume to the rendered image intensity at a given pixel is calculated as the \textit{volume} of the portion of the slice which projects into the given pixel times the intensity multiplier of the volume (Fig.~\ref{fig_mesh_unit}, right panel). This is an equivalent of the line-of-sight depth of the feature times its filling factor (to get the actual emission measure, this quantity needs to be divided over the pixel area, which is the same for all pixes and so is irrelevant in our analysis). The rendered image is then a sum of renderings of individual volumes of all loops on the image.

We assume isothermal hydrostatic atmospheres so the intensity multiplier at each slice was simply $n^2\propto e^{-2z/z_0}$. The scale height $z_0=40\mbox{ pix}\approx35\mbox{ Mm}$ \textit{is the same} for both loops. The displays show a square root of intensity for all three renderings, and the \textit{color scale is also the same} for all three plots. 

Note that in $x-y$ projection loop 1 (marked as a red triangle on all three renderings) appears brighter \textit{and} thinner than loop 2 (marked as a blue square). The opposite is true for $x-z$ projection: loop 2 appears brighter and thinner. At the same time, loop 1 \textit{does not appear to expand with height} on $x-y$ projection and loop 2 does not appear to expand in $x-z$ projection. An observer who is unaware of this effect and who wishes to study expansion of coronal loops is more likely to select loop 1 over loop 2 if looking at $x-y$ projection, simply because loop 1 stands out and appears visible from end to end; when looking at $x-z$ projection, such an observer might be more likely to select loop 2 for analysis for the same reasons, especially if there are more loops in the background. 

\begin{figure}[!hc] 
  \begin{center} 
     \includegraphics[width=9cm]{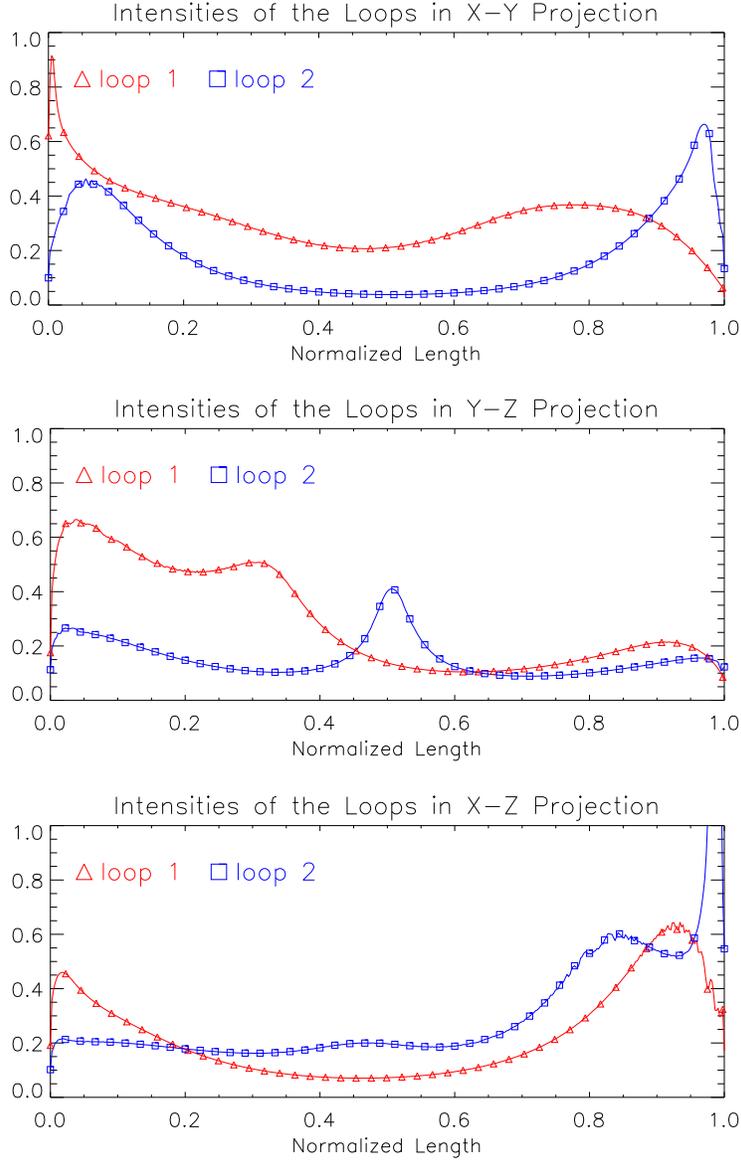} 
  \end{center} 
 \caption{Intensities of loops 1 and 2 (when rendered separately) in different projections along their axes; we took advantage of the fact that the flux tubes have similar lengths and plotted intensity along normalized length units. Note that loop 1 is brighter in $x-y$ projection at the apex while loop 2 is brighter at apex in $x-z$ projection. Note also the intensity peak at the loop top of loop 2 in $y-z$ projection. It is a real loop top feature, and we will return to it further in the text.}
 \label{2loops_intn}
\end{figure}

\begin{figure}[!hc] 
  \begin{center} 
     \includegraphics[width=9cm]{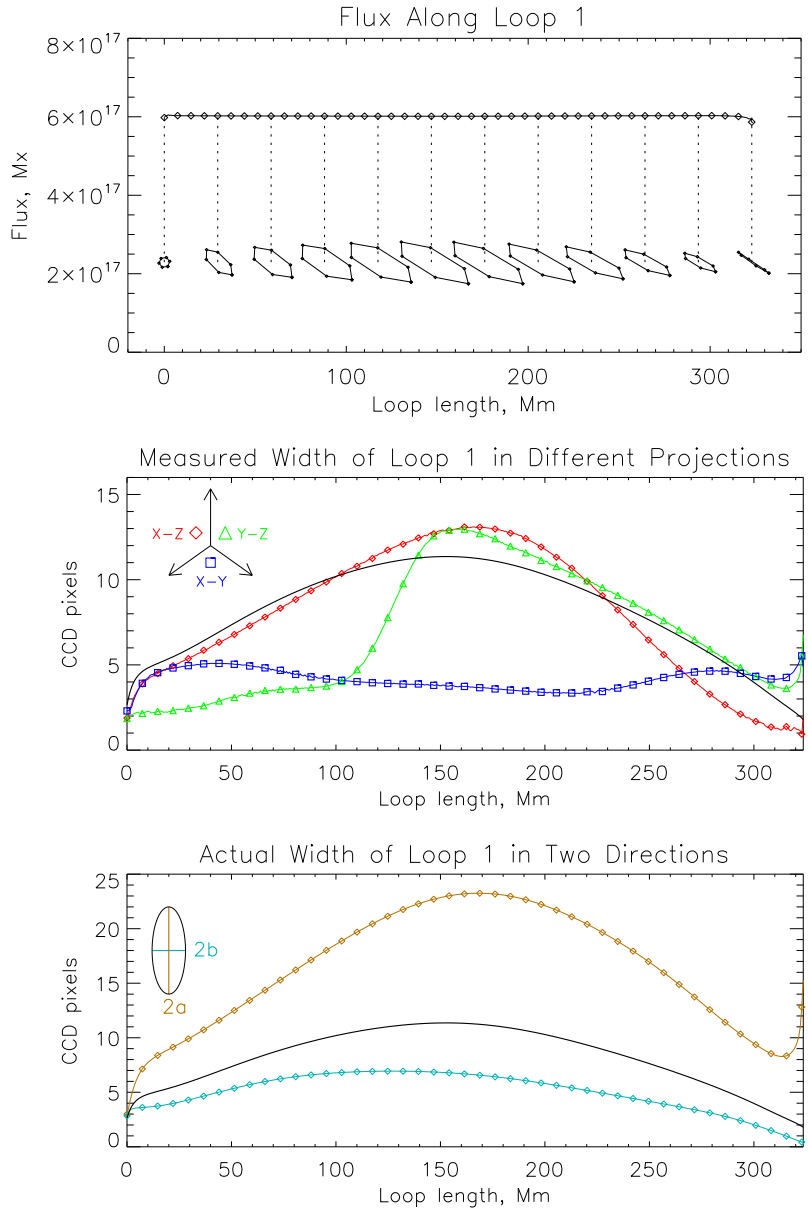} 
  \end{center} 
 \caption{Top panel shows that flux in flux tube used to render loop 1 is indeed constant; we also show several cross-sections along the tube, showing that it indeed expands and becomes squished. Middle panel shows width of loop 1, inferred from the renderings in different projections (for this, the loop was rendered without the second loop so there was no background). We took slices of the renderings along the loop, perpendicular to direction of the projected axis at each point. We then fit a gaussian profile to each slice. These inferred widths (red diamonds for $x-z$ projection, green triangles for $y-z$ and blue squares for $x-y$) are substantially different from each other and are also different from the ``expected'' width if to assume circular cross-section (solid line). Note also, that the width of the loop is plotted in pixels on the rendered images, to demonstrate that the loop is well resolved in all directions. Bottom panel shows \textit{actual} widths of loop 1, that is, major and minor axes of the ellipse fitted into cross-sections. Again, the width under ``circular cross-section'' assumption is shown for reference (black solid line).}
 \label{2loops_lp1}
\end{figure}

\begin{figure}[!hc] 
  \begin{center} 
     \includegraphics[width=9cm]{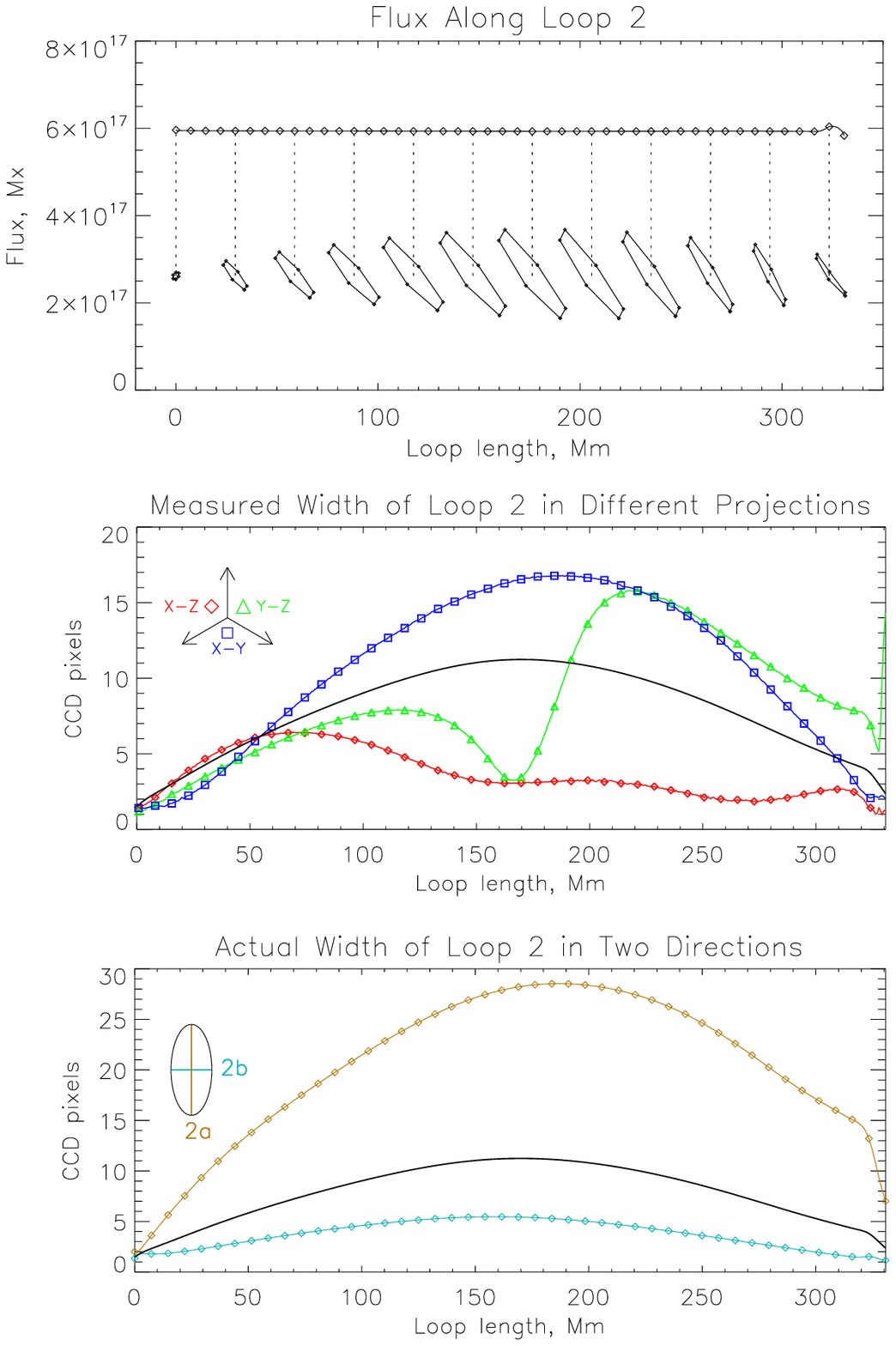} 
  \end{center} 
 \caption{Flux, cross-sections, inferred and actual expansions of loop 2 (the notation is the same as on Fig.~\ref{2loops_lp1}). Note that $x-y$ projection allows to observe the expansion of loop 2, but not loop 1, while in $x-z$ projection expansion of loop 1 is clear, but loop 2 exhibits little to none width variations. Note also that neither of the inferred widths allows to access the actual expansions of the loops.}
 \label{2loops_lp2}
\end{figure}

Further insight into the selection factor posed by the viewing angle and the implications for the inferred expansion of the loops can be obtained from Figs.~\ref{2loops_intn}-\ref{2loops_lp2}. Fig.~\ref{2loops_intn} shows \referee{computed column emission measures for each pixel} of these two loops evaluated along the axis in various projections (for this exercise, loops were rendered separately to eliminate possible background issues). It demonstrates that relative \referee{intensities} would indeed make, for example, loop 1 stand out over loop 2 at the apex in $x-y$ projection, and vice versa (though somewhat less) in $x-z$ projection. 

Figs.~\ref{2loops_lp1} and~\ref{2loops_lp2} demonstrate that, first, flux is indeed conserved in these flux tubes, second, their expansion is highly anisotropic, and third, that the actual expansion \textit{can not be measured} from any single perspective. We calculate flux along the tube as $\Phi_i=|\bvec_i|A_i$, where $A_i$ is the area of polygon forming the cross-section and $|\bvec_i|$ is evaluated at the axis; top panels on Figs.~\ref{2loops_lp1} and~\ref{2loops_lp2} show that it is indeed constant along the tubes. These panels also show the cross-sections along the tubes in several locations, demonstrating that they are substantially different from the regular starting hexagon. The middle panels show the \textit{inferred} cross-section from the three renderings (blue squares for $x-z$, green triangles for $y-z$ and red diamonds for $x-y$). We simulated the procedure of inferring width of loops from the observations \citep{Fuentes2006, Fuentes2008} in ideal circumstances: we rendered each loop separately (so there was no background) and fit gaussian profiles to slits across the projected loops' axis. The resulting widths are consistent with Fig.~\ref{2loops_3d}, for instance, loop 1 expands in $x-z$ rendering and \textit{does not expand} in $x-y$ rendering --- that is it does not expand in the projection in which it appears \textit{brighter} than loop 2 (Fig.~\ref{2loops_intn}, top panel). The units of width are chosen to be pixels on the rendered images, to demonstrate that both loops are well resolved in each projection, so the under-resolution is not an issue. For reference, we show the ``circular'' diameter of the loop under the assumption of circular cross-section, defined as $2r$, where $r=\sqrt{A/\pi}$. Note that the inferred widths do not resemble this ``circular'' diameter (except maybe for loop 1 in $x-z$ projection). The bottom panels on Figs.~\ref{2loops_lp1} and~\ref{2loops_lp2} show the \textit{actual} width of these two loops. Since the expansion is anisotropic, there are two widths shown; we fit an ellipse in each cross-section using least squares method \citep{ellipse_fit} and plot the major and minor axes $2a$ and $2b$. Again, the ``circular'' width $2r$ is plotted for reference; note that $r^2=ab$ in correspondence with the area of the ellipse being $A=\pi ab$ (there is a systematic difference so $r\approx 1.1\times \sqrt{ab}$, this is due to the difference between the area of a polygon and a circle fit to its corners).

These two loops are well within the typical range of oblateness (Table~\ref{table_stats}). Loop 1 has the aspect ratio at the apex $b_{apx}/a_{apx}\approx 5.2$ and for loop 2 $b_{apx}/a_{apx}\approx 3.4$. Yet, as we demonstrated, this oblateness was sufficient to prevent us from measuring their expansion from a given projection. 

Note also that the density scale height appears bigger for loop 2 in $x-z$ projection but the opposite is true in $x-y$ projection. The density scale heights are in fact the same for both loops, so in this example this effect is purely an artifact of loops' cross-sectional shape. We will discuss this issue further in more \refereetwo{detail}. 

\begin{figure}[!hc] 
  \begin{center} 
    \includegraphics[width=11cm]{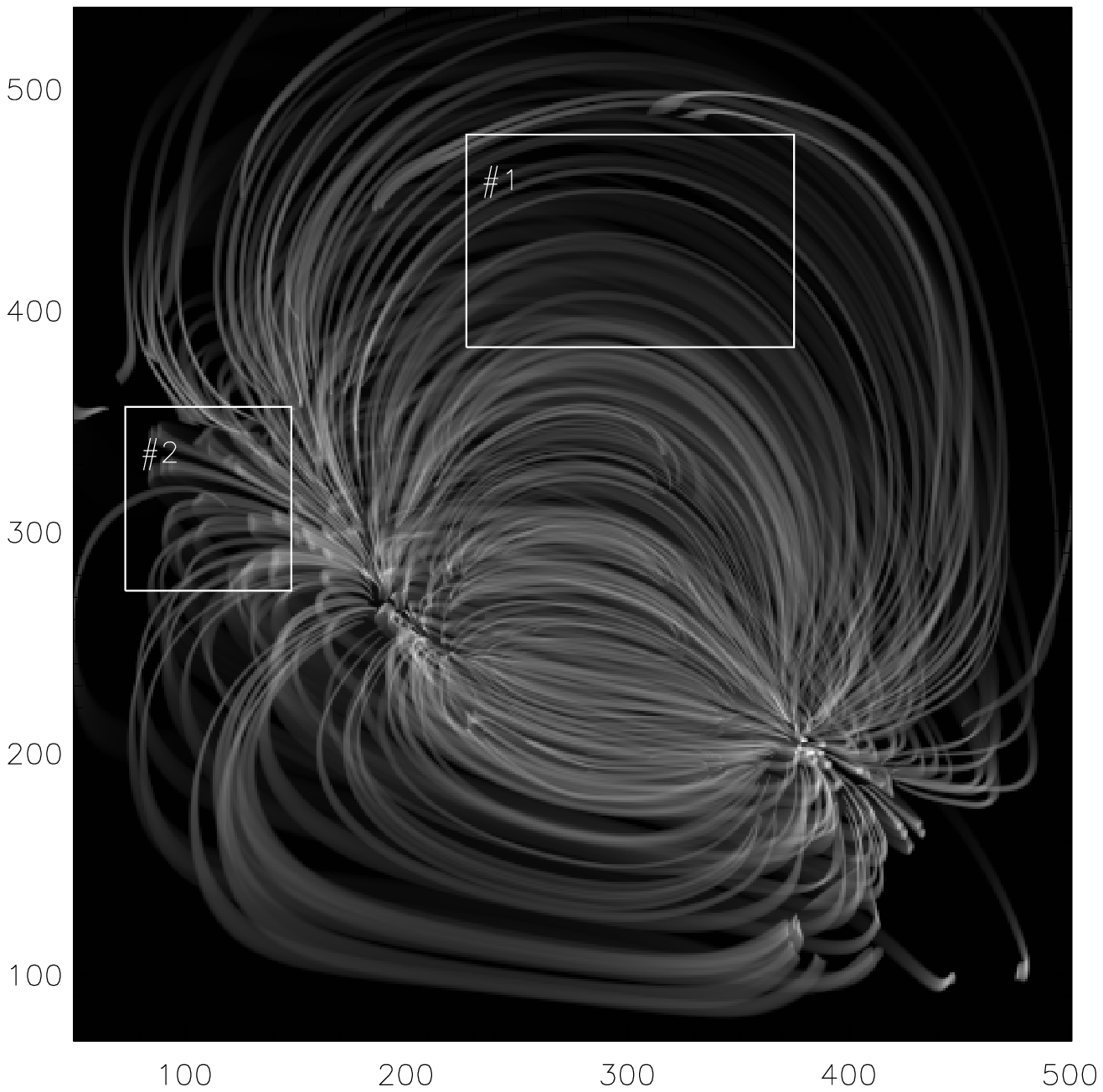} \\
    \vspace{0.75cm}\includegraphics[width=11cm]{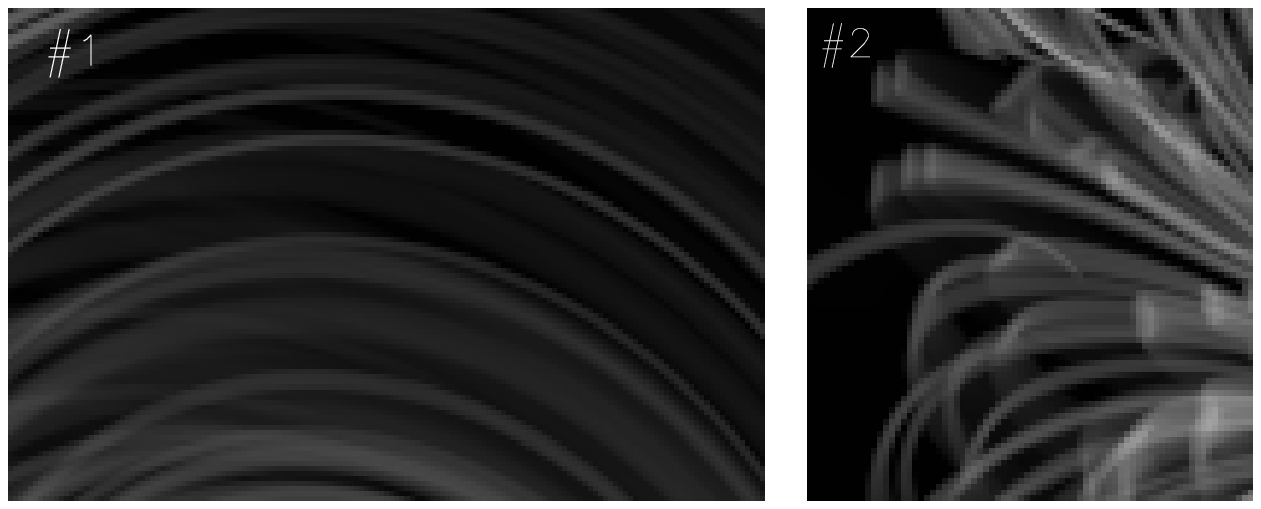} \\
    \vspace{0.5cm}\includegraphics[width=11cm]{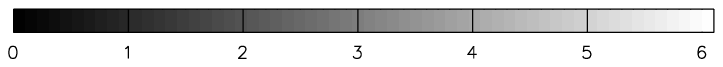} \\ 
  \end{center}
  \caption{\referee{Rendering of a set of several hundreds of flux tubes filled with hydrostatic isothermal atmospheres (with the same scale height) viewed in a plane-of-sky projection. The grayscale corresponds to square of column emission measure in arbitrary units. The flux tubes are initiated at $z=3$pix plane and the radius at the base is 0.5pix on the rendered image, so all of them are resolved at the bases. Note how tubes which do not expand a lot in the plane of the sky appear brighter and stand out from their neighbors, which do expand in this projection --- this is particularly evident on the close-up \#1. Also note characteristic elongated bright spots at where loops turn towards the observer (close-up \#2).}}
  \label{render_los}
 \end{figure}

\begin{figure}[!hc] 
  \begin{center} 
    \includegraphics[width=11cm]{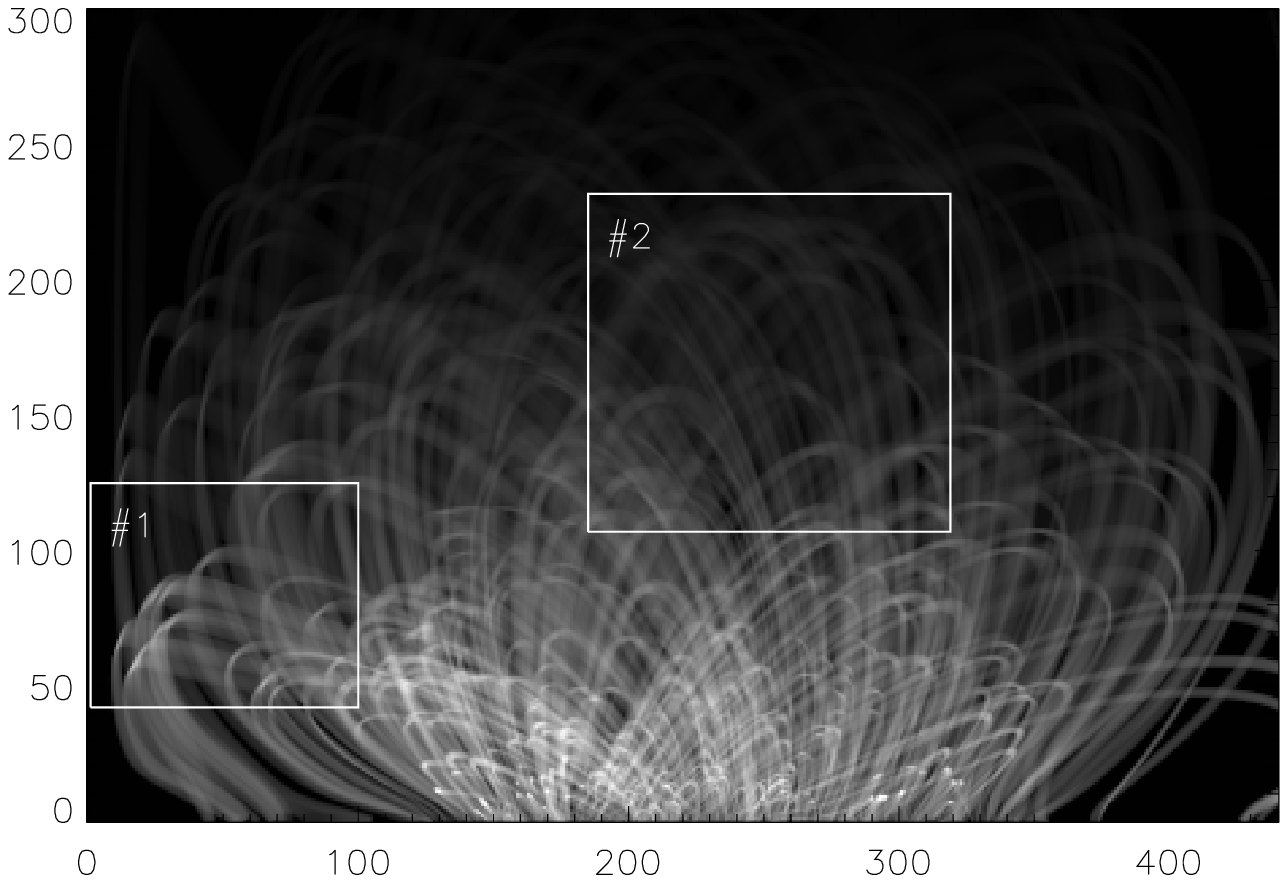} \\
    \vspace{0.75cm}\includegraphics[width=11cm]{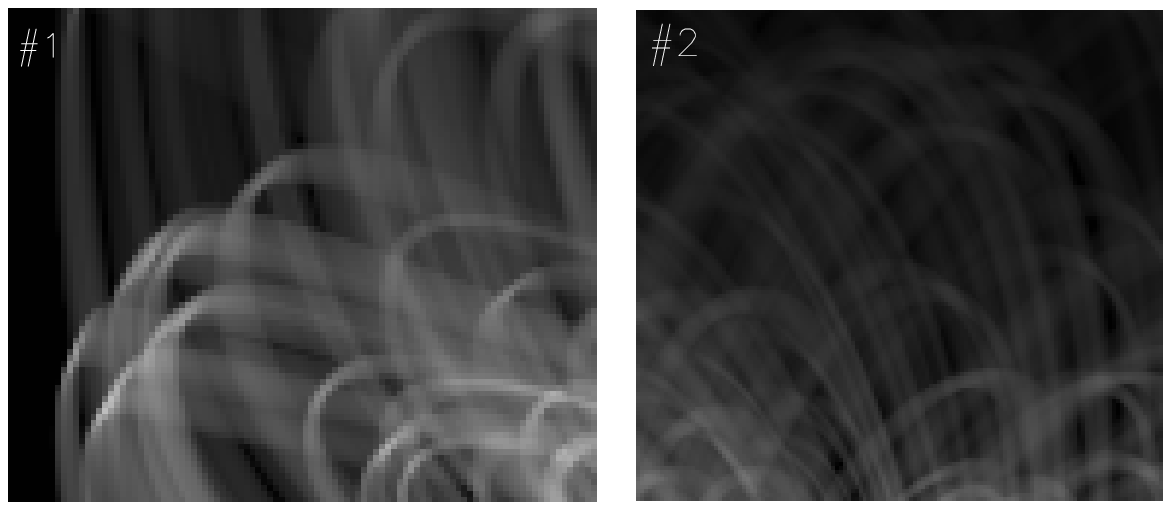} \\
    \vspace{0.5cm}\includegraphics[width=11cm]{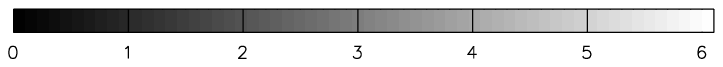} \\ 
  \end{center}
  \caption{\referee{Rendering of the same set of loops as on Fig.~\ref{render_los} but viewed from the side. Note, again, characteristic flat loop tops and elongated brightening at the loop tops (close-up \#1). Note also (close-up \#2) how selection bias highlights loops which expand less in the projection (those on the close-up \#2 which go from top left to bottom middle) against those that expand more in the projection and are narrower along line of sight (which go from bottom left to top right in the close-up frame). All of these synthetic loops have the same size and shape at the base.}}
  \label{render_side}
 \end{figure}

\begin{figure}[!hc] 
  \begin{center} 
   \begin{tabular}{cc}
     \includegraphics[height=5cm]{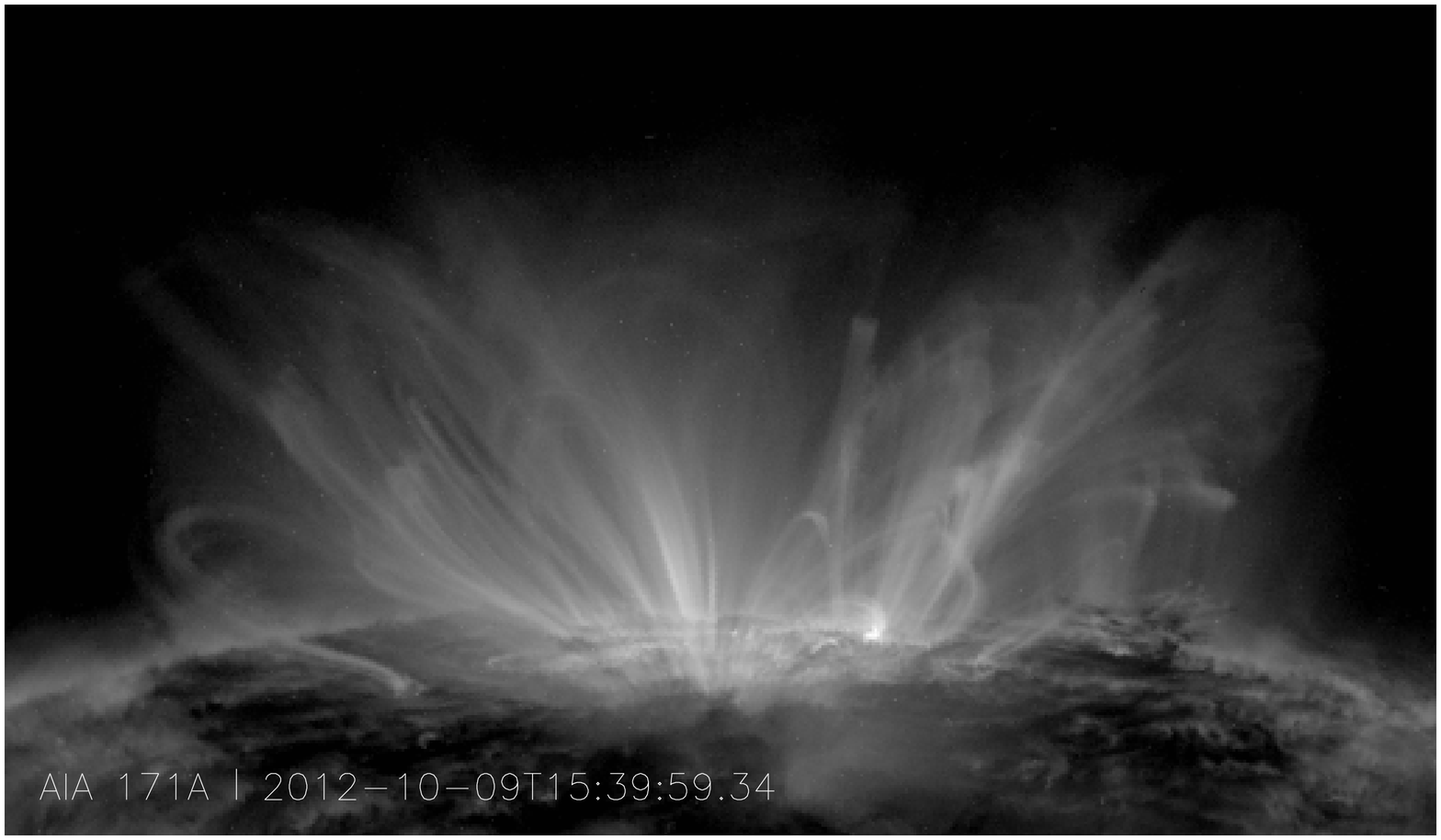} & \includegraphics[height=5cm]{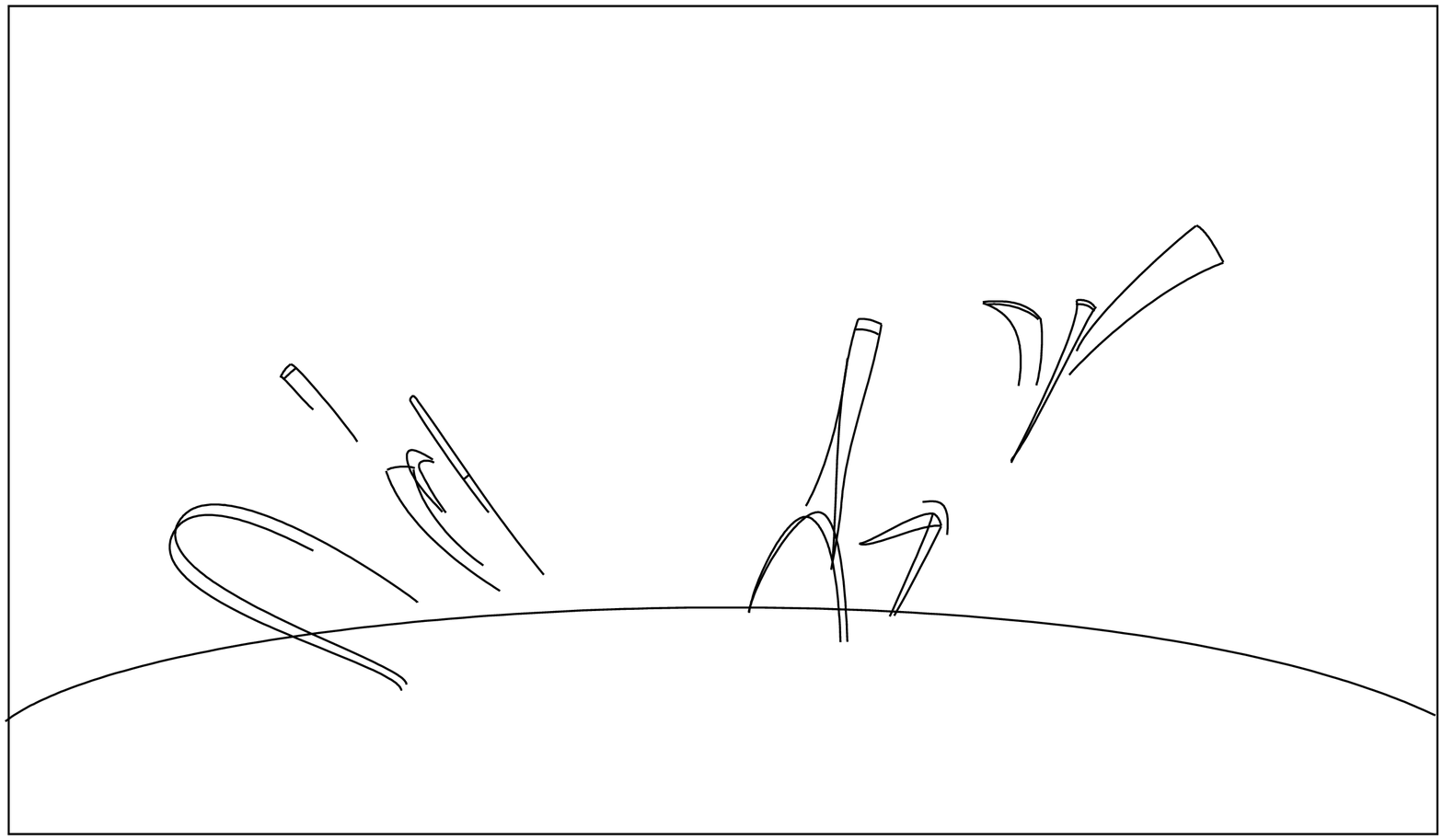} \\ 
  \end{tabular} 
 \end{center} 
 \caption{An AIA 171A image of an active region on the limb (left panel) and a schematic drawing of some features to which we wish to draw the reader's attention (right panel). Note how thickness of loops changes at \refereetwo{the apex} and at horizontally elongated brightness enhancement at loop tops. These features are also present on the synthetic images which account for realistic cross-section of flux tubes (Figs.~\ref{render_los} and~\ref{render_side}), which \referee{supports the hypothesis that cross-section of coronal loops is not in general circular}.}
 \label{aia_looptops}
\end{figure}

The renderings of many flux tubes shown on Figs.~\ref{render_los} and~\ref{render_side} confirm our findings about the lack of \textit{apparent} expansion due to the selection effect. Flux tubes which expand mostly along the line of sight appear brighter and stand out. This is particularly evident in the top middle portion of Fig.~\ref{render_los}. Note that all flux tubes were made to be about 1 image pixel wide at the base (the shown images are about 450 pixels in size), to isolate this effect from that of the unresolved strands. Fig.~\ref{render_side} also shows some loop top brightening of the shape characteristic of ribbons. Compare these loop tops to the coronal image on Fig.~\ref{aia_looptops}, left panel. The right panel of Fig.~\ref{aia_looptops} is a sketch of several other features on the coronal image which might be attributed to the oblate cross-section of the coronal loops. 

\clearpage
\section{Effect on the Observed \referee{Intensity} Scale Height}\label{sec_scale_height}

As we mentioned before, the oblateness in a loop cross section may introduce a bias in our interpretation of observations when deriving a pressure scale height. The reason for that is that to derive density, an estimate of a column depth is usually needed. As we showed in Section~\ref{sec_2loops}, this column depth does not in general follow from the width of the loop. 

\begin{figure}[!hc] 
  \begin{center} 
   \begin{tabular}{c|c|c|cc|c}
     Expansion & 1. None & 2. Isotropic & 3. Anisotropic & 4. Anisotropic & 5. Unresolved \\ 
     & & & $\bot$ to the l.o.s. & along l.o.s. & $\bot$ to the l.o.s. \\ 
     & & & & & \\
     & \includegraphics[height=2cm]{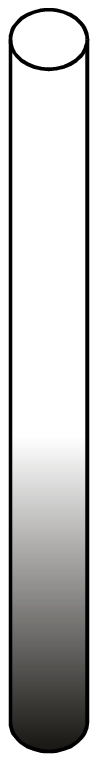} & 
     \includegraphics[height=2cm]{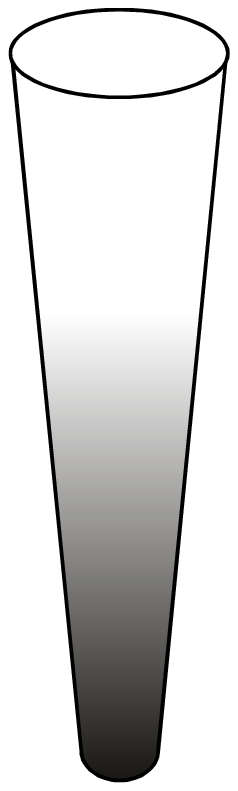} & 
     \includegraphics[height=2cm]{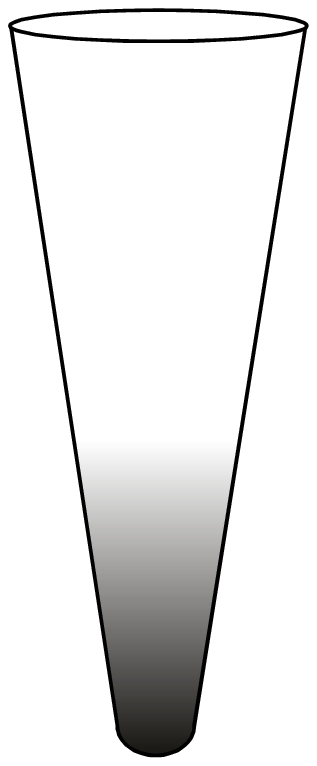} & 
     \includegraphics[height=2cm]{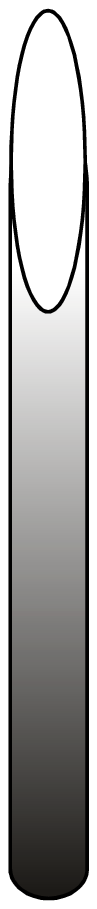} &      
     \includegraphics[height=2cm]{f035.eps} \\     
     & & & & & \\
     Flux & irrelevant & $\Phi\propto Bd^2$ & \multicolumn{2}{c|}{$\Phi\propto Bab$, $b=const$} & $\Phi=BA$\\
     Column depth & $d(z)=const$ & $d(z)\propto B^{-1/2}$ & $d(z)=const$ & $d(z)\propto B^{-1}$ & $d_{\mbox{eff}}\propto B^{-1}$ \\
     $EM(z)\propto$ & $n^2$ & $n^2B^{-1/2}$ & $n^2$ & $n^2B^{-1}$ & $n^2B^{-1}$ \\
  \end{tabular} 
 \end{center}
 \caption{Impact of the expansion of coronal loops on the \textit{observed} density scale height.}
 \label{scale_height_sketch} 
\end{figure}

To see this, consider a vertical flux tube filled with hydrostatic isothermal atmosphere. When the tube is viewed from the side, its emission measure \referee{at a given image pixel along its projected axis} will be $EM(z)\propto n^2d$, where $d$ is the column depth. If the diameter of the flux tube is constant with height \citep[e.g., as in][]{Rosner1978}, as in Fig.~\ref{scale_height_sketch}, column 1, then $EM(z)\propto n^2(z)$. Flux tube expansion with height increases the geometrical column depth and therefore $EM(z)$ drops slower given the same pressure scale height (Fig.~\ref{scale_height_sketch}, column 2). Flux $\Phi$ is constant along the tube, so if the cross-section is circular, then the column depth is $d(z)=2\sqrt{\Phi/\pi B(z)}$ and in this case $EM(z)\propto n^2B^{-1/2}$. Now suppose that $B(z)$ drops with height but that the flux tube expands only in one direction and that its cross-section is an ellipse with semiaxes $a>b$ and area $\pi ab$. In this case, when viewed \textit{perpendicular} to the expanding direction, the column depth would stay constant and $EM(z)\propto n^2$ like for a non-expanding tube. When viewed along the direction of expansion, \textit{the same flux tube} would fade as $EM(z)\propto n^2 B^{-1}$ (Fig.~\ref{scale_height_sketch}, columns 3 and 4). In unresolved case the tube is thinner than the pixel width $\Delta x_{pix}$ in the plane of sky, as in column 5 on Fig.~\ref{scale_height_sketch}, its emission times its filling factor would be equivalent to emission of a tube one pixel wide with depth $d_{eff}=A/\Delta x_{pix}$. In this case $EM(z)\propto n^2B^{-1}$ regardless of the cross-section shape, and is the same as in the case of \textit{resolved} tube expanding along line of sight only. 

\begin{figure}[!hc] 
  \begin{center} 
   \includegraphics[height=6cm]{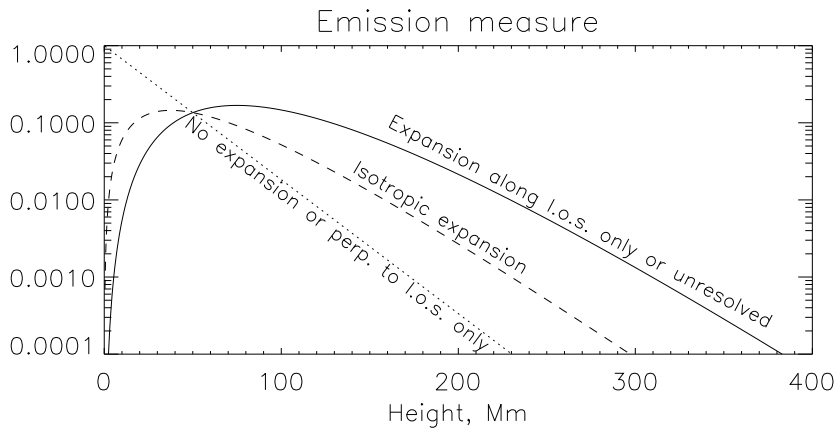} \\ 
  \end{center}
  \caption{Model of how emission measure scales with height for the flux tubes on Fig.~\ref{scale_height_sketch}. The functions plotted are $n^2$ (dotted line), $n^2B^{-1/2}$ (dashed line) and $n^2B^{-1}$ (solid line). We assume hydrostatic atmosphere $n\propto \exp(-z/z_0)$ with $z_0=50$Mm for all curves and the field strength dropping like that of a point dipole, $B(z)=z_0/z^3$ (the constant for $B$ is chosen to be such that all three curves intersect at one density scale height).}
  \label{scale_height_plot}
 \end{figure}

Fig.~\ref{scale_height_plot} shows these three power laws, $n^2$, $n^2B^{-1/2}$ and $n^2B^{-1}$, plotted against $z$ \citep[essentially revisiting Fig. 5 in][]{DeForest2007}. All three models assume $n=exp(-z/z_0)$ with $z_0=50$Mm and $B\propto 1/z^3$, with the coefficient chosen such that all three curves intersect at one pressure scale height. Both unresolved and l.o.s. only expansion ($EM\propto n^2B^{-1}$) can be an order of magnitude brighter than the constant cross-section curve ($EM\propto n^2$), which is in agreement with the observations \citep{Warren2003b}, for example, on our plot at two pressure scale heights. The anisotropic expansion model; however, is supported by the study of \citet{Fuentes2008} who showed that it is possible to distinguish between expanding, expanding but unresolved, and resolved but not expanding loops. \referee{We also demonstrate, that, depending on orientation, the scale height might even be \textit{lower} than that of an isotropically expanding loop. In fact, the same fan-like structure, which is not entirely planar but has ``ripples'' on it, might appear in the plane of sky as a combination of dim underdense diffuse emission and thin overdense bright strands --- despite having uniform pressure scale height!}

\section{Summary and Discussion}\label{sec_discussion}

We have demonstrated that magnetic flux tubes \referee{may} exhibit a lot of variation in cross-sectional shape along their length. In particular, flux tubes with round bases at the lower boundary become oblate at their apices \referee{in the potential field models that we examined}. The direction and the amount of oblateness varies greatly between different field models and even within a single model. This in principle might render a study of cross-sectional shape based on \textit{one} selected flux tube inconclusive. Therefore, analysis of large statistical samples of flux tubes is required to draw definite conclusions on their shape. 

\referee{We examined a large sample of thin flux tubes in a potential field model of an active region.} The conclusion is that at least in \referee{the studied model} the aspect ratio at the apices of initially (at the base) round tubes varies in around 1.5 to 5. The distribution peaks at 2 but it is highly skewed and has the median at around 4.5. We verified that the oblateness is only slightly influenced by the inclination of the field at the base and is therefore not strongly modulated by the initial cross-sectional shape of the flux tube. \referee{We also examined the shape distortions for the end-to-end mapping of the flux tubes and concluded that, in general \refereetwo{in} our model field, these distortions are even stronger than the end-to-apex ones. That in principle implies that for a given coronal loop, two footpoints might be of substantially different shape, which in turn could affect heating models which assume that heat is deposited on both footpoints. This; however, goes outside of the scope of this paper; we chose to focus the paper on the apex properties of coronal loops.}

These values of oblateness can strongly influence the perceived expansion of flux tubes, making it depend on the viewing angle. In particular, we demonstrate for two flux tubes, drawn from the model field, that viewed from the top, one of the loops expands and the other does not, whereas the opposite is seen when viewed from the side. 

We address the arguments of \citet{Klimchuk2000} who concluded that coronal loops must be generally circular in cross-section. The flux tubes in our rendering exercise, while being strongly oblate in cross-section, produced simple, single peak profiles on the rendered images. These profiles could be fit with gaussian curves and yield widths in the manner similar to what is currently done in the literature \citep[e.g.,][]{Fuentes2008}. 

\referee{We demonstrated that the widths, inferred from a simulated image, do not correlate with the actual width of these flux tubes. Neither do they correlate to diameters of circles of the same area as the flux tubes. In the two projections in which the flux tubes are well visible, their widths vary rather modestly, again, despite their strong oblateness.} 

We also make a point that for oblate optically thin flux tubes, variations in width are associated with variations in brightness: when viewed along the ``wide'' side, a ribbon appears thin \textit{and} bright, and when viewed perpendicular to it, the same ribbon appears wide \textit{but} dim. This also addresses the argument of \citet{Fuentes2006} who pointed that even if loops were oblate, a spread in orientations with respect to the line of sight would still allow us to observe their expansion --- on average. It is an important point of this paper, as it implies that for any given viewing angle, there will be \textit{selection criteria} which will make some ribbons stand out when they are viewed along the ``wide'' side. 

\referee{Our finding also addresses the increased pressure scale height observed in some bright EUV loops \citep[e.g.,][and references therein]{Warren2003b, Winebarger2003} at the same time being consistent with evidence that individual strands are in fact resolved \citep{Fuentes2008, Aschwanden2005}.} As we showed in the previous section, a resolved but highly oblate loop might demonstrate the same enhancement of apparent scale height as would an unresolved loop.  

\begin{figure}[!hc] 
  \begin{center} 
   \includegraphics[height=6cm]{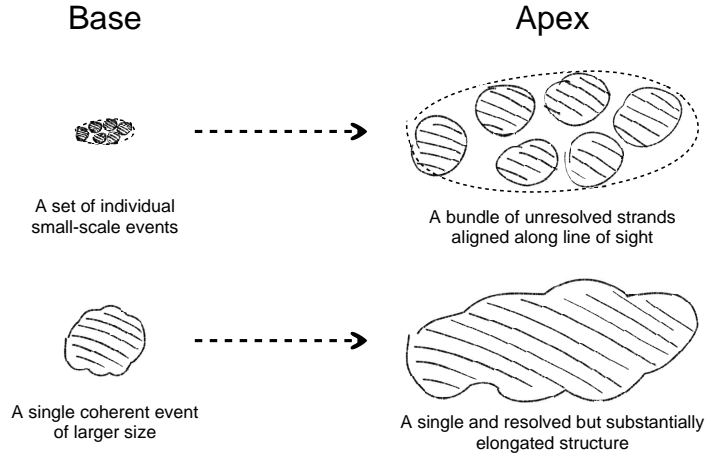} \\ 
  \end{center}
  \caption{Without prior knowledge of how magnetic flux tubes are distorted at a particular location, it would be hard to tell apart the emission produced by the two scenarios sketched in this figure. The first one is a result of several small-scale events which supply the corona with hot plasma; they are small enough to stay unresolved (or barely resolved) at the apex, and the plasma they supply is dense enough so that the structure is visible even with the low filling factor. The second one is a result of a single coherent event, but the flux tube which embeds this event gets elongated at the apex; such event might be not much smaller than perceived width of the structure at the apex. As the edges of the flux tube might not, in general, be smooth, the fluctuations in line of sight depth might produce small ``ripples'' in otherwise resolved structure.}
  \label{bundle_vs_single_structure}
 \end{figure}

\begin{figure}[!hc] 
  \begin{center} 
   \includegraphics[height=8cm]{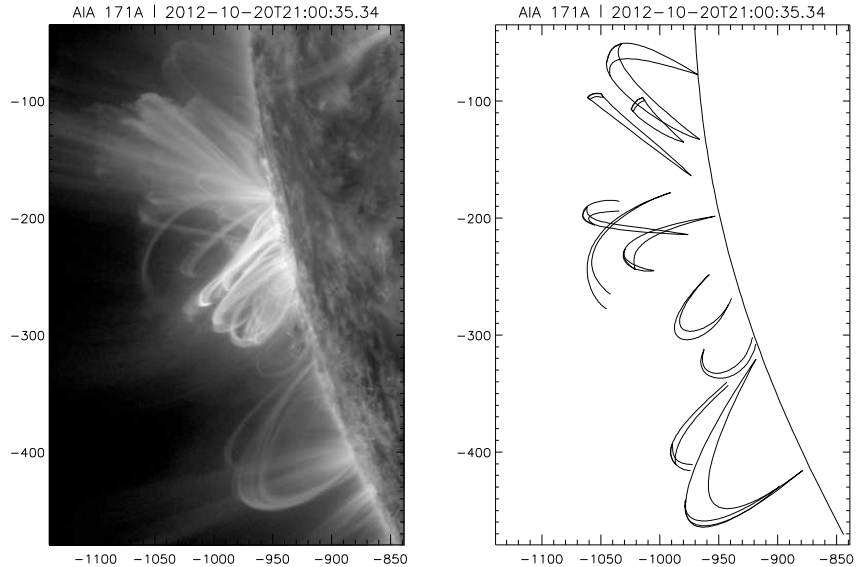} 
  \end{center}
  \caption{An AIA 171\AA~image and a sketch of several features we wish to draw attention to. In the light of Fig.~\ref{bundle_vs_single_structure}, these features might be composed of bundles of thin strands, or they might be ribbons of non-uniform thickness. Note in particular how loops become dimmer as they get wider, even at the same height above the surface. This is an evidence of their oblate cross-sectional shape. Also note the flat loop tops, which are also consistent with the the hypothesis of their oblate cross-section.}
  \label{bundle_vs_single_structure_p2}
 \end{figure}

Immediate implications of our study are the following: without analysis of shape distortions between the bases and the apices of coronal loops we cannot \referee{put constraints on} the size of the loop base, nor can we \referee{evaluate by how much is plasma in the loops denser} compared to the diffuse background. For example (see Fig.~\ref{bundle_vs_single_structure}), the same non-expanding coronal loop could be composed of a set of thin unresolved strands with small bases (so that they are unresolved at the apex) and high density (to compensate for the low filling factor), or it could be a \textit{single coherent \referee{structure}} of a larger size at the base, which does not expand along the line of sight. \referee{Moreover, if the cross-section is not a smooth shape but rather has a ``wrinkled'' edge \citep[e.g.,][reported that in their model field, circular cross-sections of thin flux tubes deformed into rippled shapes further along the axes]{Gudiksen2005b}, the individual ``wrinkles'' may, due to the increased column depth, appear brighter. Therefore viewed at higher resolution, both systems might resolve into thin strands.} Fig.~\ref{bundle_vs_single_structure_p2} points at such features on the Sun. Even if to consider that some (or all) of these are bundles of strands, these bundles appear to be elongated in cross-section, which by itself is consistent with our \referee{finding}. 

\referee{Our study is focused on the properties of magnetic \textit{flux tubes}; however care must be taken to distinguish between them and observed \textit{coronal loops}. Loops fill only a fraction of the coronal volume and the selection mechanisms which determine which flux tubes get filled with hot and dense plasma are a topic of ongoing research. It might in principle be possible that, for example, plasma is more likely to get heated in flux tubes which exhibit a relatively isotropic expansion (in our experiment, as Fig.~\ref{apex_stats_ab} shows, out of roughly 9000 flux tubes, there is about a thousand for which $a_{apx}/b_{apx}\leq 2$). In that case, coronal \textit{loops} might predominantly have a circular cross-section even if for \textit{flux tubes} this is not generally true. An evidence against that is that loop fans are associated with topological divides and therefore with regions of high, rather than low, distortions \citep{Schrijver2010}. It might also be the case that the events which result in the deposition of hot plasma into the atmosphere happen predominantly at coronal heights, and are approximately round across the field --- in the corona. Our preliminary experiments suggest that most of the shape distortion happens relatively close to the lower boundary (e.g., see cross-sections of our simulated loops on Figs.~\ref{2loops_lp1} and~\ref{2loops_lp2}), so flux tubes round at their apices might undergo less shape distortion at their higher portions, in which case they would have highly distorted footpoints. If observational evidence in favor of either scenario become available, that might be very important for our understanding of the mechanism of coronal heating. In the absence of such, our analysis gives us reasons to believe that loops should not, in general, be assumed to have circular cross-section through their length.}

\referee{Overall, our study allows to explain constant width of coronal loops as well as their excessive pressure scale height using fewer assumptions. Indeed, rather than involving additional entities in the explanation (such as size of the strands, coronal currents and so on), we lift the assumption of the circular cross-section. We show that the aforementioned properties of loops might be a natural consequence of a simple property of magnetic field that has not been analyzed to date.}

\bibliography{c:/localtexmf/bib/short_abbrevs,c:/localtexmf/bib/full_lib,c:/localtexmf/bib/my_bib}
\end{document}